\def\eps@scaling{1.0}

\newcommand\plotone[1]{%
 \centering
 \leavevmode
 \includegraphics[width={\eps@scaling\columnwidth}]{#1}%
}% 

\makeatletter
\def\itemize{%                                                                   
  \ifnum \@itemdepth >\thr@@\@toodeep\else
    \advance\@itemdepth\@ne
    \edef\@itemitem{labelitem\romannumeral\the\@itemdepth}%                      
    \expandafter
    \list
      \csname\@itemitem\endcsname
      {\itemsep=0pt\parsep=0pt\topsep=0pt\def\makelabel##1{\hss\llap{##1}}}%     
  \fi}
\makeatother
\documentclass[letterpaper,11pt]{article}
\usepackage{graphicx,times,amssymb,dashrule,natbib,aas_macros,url,hyperref,enumerate}
\usepackage{arydshln}
\usepackage{color}
\usepackage{hyperref}
\usepackage[section]{placeins}
\usepackage{pdfpages}
\usepackage{comment}
\newcommand{\smid}{SKA1-mid\,\,}
\newcommand{\low}{SKA1-low\,\,}
\hypersetup{
    colorlinks,
    citecolor=blue,
    filecolor=green,
    linkcolor=red,
    urlcolor=blue
}
\makeatletter
\renewcommand{\section}{\@startsection%
{section}{1}{0mm}{-\baselineskip}%
{0.5\baselineskip}{\normalfont\Large\bfseries}}%
\makeatother
\begin{document}
\pagestyle{plain}
\pagenumbering{arabic}
%%%%%%%%%%%%%%%%%%%%%%%%%%%%%
%%%%% Title of proposal %%%%% 
%%%%%%%%%%%%%%%%%%%%%%%%%%%%%
\title{Clusters of galaxies and the cosmic web with SKA}
\author{Ruta Kale$^1$
  \texttt{(ruta\,AT\,ncra\,DOT\,tifr\,DOT\,res\,DOT\,in)}, \\
 K. S. Dwarakanath$^{2}$
  \texttt{(dwaraka\,AT\,rri\,DOT\,res\,DOT\,in)},\\
 Dharam Vir Lal$^1$, Joydeep Bagchi$^{3}$, Surajit Paul$^{4}$, Siddharth Malu$^{5}$, \\
 Abhirup Datta$^5$, Viral Parekh$^2$,  
 Prateek Sharma$^{6}$ and Mamta Pandey-Pommier$^{7}$\\\\ 
$^1$ National Centre for Radio Astrophysics, T. I. F. R., Pune 411007, India\\
 $^{2}$ Raman Research Institute, C. V. Raman Avenue, Sadashivanagar, Bangalore 560080, India\\
 $^{3}$ Inter University Centre for Astronomy and Astrophysics, Post Bag 4, Pune 411007, India\\
 $^{4}$ Department of Physics, Savitribai Phule Pune University, Pune 411007, India\\
 $^{5}$ Indian Institute of Technology, Khandwa Road, Simrol, Indore 453552, India\\
 $^{6}$ Indian Institute of Science, C. V. Raman Avenue, Bangalore 560012, India\\
  $^{7}$ Univ Lyon, Univ Lyon1, Ens de Lyon, CNRS, Centre de Recherche Astrophysique de Lyon UMR5574, \\
  F- 69230, Saint-Genis-Laval, France\\}
\date{}
\maketitle
\tableofcontents
%\listoffigures
%\listoftables
%\pagebreak
%
%\section*{List of Contributors}
%To be added.
%
%\section{Abstract}

\section*{Abstract}
The intra-cluster and inter-galactic media that pervade the large scale structure of the Universe 
are known to be magnetised at sub-micro Gauss to micro Gauss levels and to contain cosmic rays.
The acceleration of cosmic rays and their evolution along with that of magnetic fields in these media is
still not well understood.
Diffuse radio sources of synchrotron origin associated with the intra-cluster medium 
(ICM) such as radio halos, relics and mini-halos
are direct probes of the underlying mechanisms of cosmic ray acceleration.
{Observations with radio telescopes such as the Giant Metrewave Radio Telescope,
the Very Large Array and the Westerbork Synthesis Radio Telescope have led to the discoveries of about
80 such sources and allowed detailed studies in the frequency range 0.15 - 1.4 GHz of a few.}
These studies have revealed scaling relations between the thermal {and non-thermal} properties of clusters and
favour the role of shocks in the formation of radio relics and of turbulent re-acceleration in the formation of 
 radio halos and mini-halos.
The radio halos are known to occur in merging clusters and mini-halos are detected in about half of the 
cool-core clusters.
Due to the limitations of current radio telescopes, low mass galaxy clusters and galaxy groups
remain unexplored as they are expected to contain
much weaker radio sources. Distinguishing between the primary and the secondary models of  
cosmic ray acceleration mechanisms  requires spectral measurements
over a wide range of radio frequencies and {with high} sensitivity. Simulations have also
predicted weak diffuse radio sources associated with filaments connecting galaxy
clusters.

The Square Kilometer Array (SKA) is a next generation radio telescope that will operate in 
the frequency range of 0.05 - 20 GHz with unprecedented sensitivities and resolutions. 
The expected detection limits of SKA will reveal a few hundred to thousand new radio halos, relics and
mini-halos providing the first large and comprehensive samples for their study. The wide frequency coverage
along with sensitivity  to extended structures will be able to constrain the cosmic ray acceleration mechanisms.
The higher frequency  ($>5$ GHz) observations will be able to use the Sunyaev-Zel'dovich effect to probe the ICM 
pressure in addition to the tracers such as lobes of head-tail radio sources.
The SKA also opens prospects to
detect the ``off-state'' or the lowest level of
radio emission from the ICM predicted by the hadronic models and the turbulent re-acceleration models.

\section{Overview}
The Square Kilometer Array (SKA) is the next generation radio telescope  
that will probe the fundamental physics of formation and evolution of galaxies up to 
large scale structures in the Universe. 
The SKA has a low frequency component (\low) that will be built in Australia and 
a high frequency component (\smid) to be built in South Africa. 
This document gives an overview of the scientific interests of the Continuum Science Working Group 
members of SKA India in the field of galaxy clusters and the cosmic web. 

\section{Current SKA1 parameters}
A brief description of the currently planned SKA1-mid and the SKA1-low telescopes is 
provided below. The complete details can be 
found in the SKA document 
released in November 2015 
\footnote{SKA-TEL-SKO-0000308\_SKA1\_System\_Baseline\_v2\_DescriptionRev01-part-1-signed.pdf. 
Note that the ``SKA1-SUR'' for which predictions can be found in 
the articles in AASKA 2014 book has been deferred.}.

\subsection{SKA1-mid (0.35 - 13.6 (20) GHz)}
The SKA1-Mid telescope is proposed to be a mixed array of 133 15-m SKA1 dishes and 64 13.5-m 
diameter dishes from the Meer Karoo Array Telescope (MeerKAT). The antennas will be arranged in a moderately compact core with a 
diameter of $\sim$1 km, a further 2-dimensional array of randomly placed dishes out to 
$\sim$3 km radius, thinning at the edges. Three spiral arms will extend to a radius of 
$\sim$80 km from the centre.\\
The dishes will be capable of operations up to at least 20 GHz, although initially equipped to 
observe only up to 13.8 GHz for SKA1. MeerKAT dishes are expected to be equipped with a front-end 
equivalent to SKA Band 2 (0.95 - 1.76 GHz), a Ultra High Frequency front-end that overlaps with Band 1 (0.3 - 1 GHz), 
and an X-band front-end (8 - 14.5 GHz).
Band 2 (0.95 - 1.76 GHz), 5 (4.6 - 13.8 GHz) and 1 (0.35 - 1.05 GHz) will be constructed in priority order as written 
and the sensitivity at a fiducial frequency of 1.67 GHz is given in Table ~\ref{skasens}.

\subsection{SKA1-low (0.05 - 0.35 GHz)}
SKA1-low telescope receptors will consist of an array of $\sim$131,000 log-periodic 
dual-polarised antenna elements. Many of the antennas will be arranged in a very compact 
configuration (the `core') with a diameter of $\sim$1 km, the rest of the elements will 
be arranged in stations, each a few 10s of metres in diameter. The stations will be distributed 
over a 40-km radius region lying within Boolardy Station, most likely organised into 
spiral arms with a high degree of randomisation.\\
The antenna array will operate from 50 MHz to $\sim$350 MHz.\\
The antenna elements will be grouped into $\sim$512 stations, whose antennas will 
be beam-formed to expose a field-of-view of $\sim$20 deg$^{2}$. The expected sensitivity at 
a fiducial frequency of 0.11 GHz is given in Table ~\ref{skasens}.

%\subsection{Surface brightness sensitivity of the SKA}

\begin{table}[t]
\centering
\caption{Sensitivities of SKA1 as of Nov. 2015.}\label{skasens}
\begin{tabular}{lcccccc}
\noalign{\smallskip} \hline
\hline \noalign{\smallskip}
 &$\nu^{\dag}$& BW&rms&$\theta_b$ \\
 &GHz &MHz&$\mu$Jy b$^{-1}$ hr$^{-1/2}$& arcsec\\
 \hline \noalign{\smallskip}
\smid &1.67&770&0.75&0.25\\
%    &&0.1&66\\
%   &4.6-13.8&??&?? \\
    \hline \noalign{\smallskip}
\low &0.11&300&3.36&7\\
%    &&0.1&184&&\\
 \hline
 \hline
\end{tabular}\\
$^\dag$ Fiducial frequencies as provided by the SKA are used here.
\end{table}

\section {Introduction: Galaxy clusters, groups and superclusters}
\label{gal-cluster}
The theorized large scale structure of the Universe evolving through initial density fluctuations
\citep[e.g.][]{1970A&A.....5...84Z} has been observed to be a 
network of filaments and sheets of matter \citep{2003ApJS..148..175S}, 
interwoven like a `web' \citep{1996Natur.380..603B} and has been reproduced by cosmological 
simulations \citep[e.g.][]{2005Natur.435..629S}. 
Galaxy clusters are the nodes of the `web' having typical masses $\sim10^{14-15}$ M$_{\odot}$ and 
 containing hundreds to thousands of galaxies moving with velocities $\sim700 - 1000$ km s$^{-1}$  
 \citep[e.g.][]{1993ApJ...404...38G}. 
Diffuse thermal gas of temperature $\sim10^{7-8}$ K (a few keV) at cluster cores 
emits thermal Bremsstrahlung making clusters extended, soft X-ray emitting sources 
\citep[e.g.][]{1966ApJ...146..955F,
1977MNRAS.178P..75M}. This thermal plasma also contains relativistic particles (GeVs) and magnetic 
fields ($\sim
0.1 - 5 \mu$G) and is collectively 
termed as the intra-cluster medium (ICM).
Smaller, less massive ($<10^{14}$ M$_{\odot}$) bound systems of galaxies 
that are found around galaxy clusters and in filaments are termed as galaxy groups 
\citep[e.g. 3C449 and 3C288][]{2013ApJ...764...83L,2010ApJ...722.1735L}.
Loosely bound complexes of a number of galaxy clusters and groups are termed as 
superclusters.

\subsection{Generation and evolution of cosmic rays}
The detectable radio emission from extragalactic sources in and around galaxy clusters is mainly of synchrotron 
origin. Therefore it is a direct probe of processes that govern the generation and evolution of relativistic 
particles and magnetic fields in the galaxies themselves and in the diffuse media surrounding them. 
The variety of radio sources in galaxy clusters can be classified into two broad categories:
\begin{enumerate}[i.] 
  \item those associated with individual galaxies in the cluster, and, 
  \item those associated with the ICM.
\end{enumerate}
 %\item those associated with individual galaxies in the cluster and 
 %\item those associated with the ICM itself and not with individual galaxies.
%\end{enumerate}

Starbursts (supernovae), active galactic nuclei (AGN) and radio galaxies 
belong to class (i). These can be compact radio sources or extended sources, but showing obvious 
association 
with individual galaxies in the cluster.
The class (ii) sources are diffuse extended sources with sizes typically $\gtrsim100$ kpc and show no obvious 
association with individual galaxies. 
In this work we will focus mainly on class (ii) sources that are probes of the cosmic ray content and magnetic 
 field in the ICM.

 The cosmic rays once produced, will lose energy primarily by synchrotron emission and inverse 
 Compton (IC) scattering 
 of the Cosmic Microwave Background (CMB) photons.
  The synchrotron losses depend on the energy 
 and magnetic field whereas the IC-losses depend on the energy density of the CMB photons and thus,
 will scale with redshift as $(1+z)^{4}$. Relativistic protons of energy 1GeV -1 TeV at cluster cores have 
 radiative lifetimes of   
 several Gyrs but that of the relativistic electrons are about 0.01 - 0.1 Gyrs.
 The distances over which the cosmic rays can travel within 
 their radiative lifetimes is important to understand the observed source sizes.
 The dispersal of cosmic rays from their source into the ICM depends on 
 the diffusivity of the ICM. The diffusion time, $\tau_{diff}$ taken by cosmic rays to diffuse to distances 
 of Mpc assuming an optimistic spatial diffusion coefficient is several Gyrs \citep{bru14}. This implies that 
 cosmic rays once produced will be confined to the cluster and need to be produced $\it in-situ$ in the ICM in the case 
 of sources of sizes of 100s of kpc.
 
  The primary models for the in-situ 
  generation of cosmic ray electrons (CRes) are based on (re-)acceleration of electrons via shocks and turbulence and the 
  secondary models are based on hadronic collisions \citep[see][for a review]{bru14}. The secondary models 
  predict a population of cosmic ray protons (CRp) to accumulate in clusters over its formation that results in 
  relativistic electrons via the collisions of CRps and thermal protons \citep[e.g.][]{den80,bla99}. 
  An associated gamma ray flux is 
  expected but has not been detected so far; stringent upper limits exist based on the Fermi Gamma Ray observatory 
  \citep[e.g.][]{ack10,bru12,2014ApJ...787...18A}. In the primary models, shocks in the ICM and turbulence are 
  the drivers behind the acceleration of particles and thus are invoked in merging galaxy clusters.
 Radio observations provide the means to distinguish between the roles of these processes in the 
 generation of synchrotron emission from the ICM. 
  
\begin{figure}
\centering
  \includegraphics[height=6 cm]{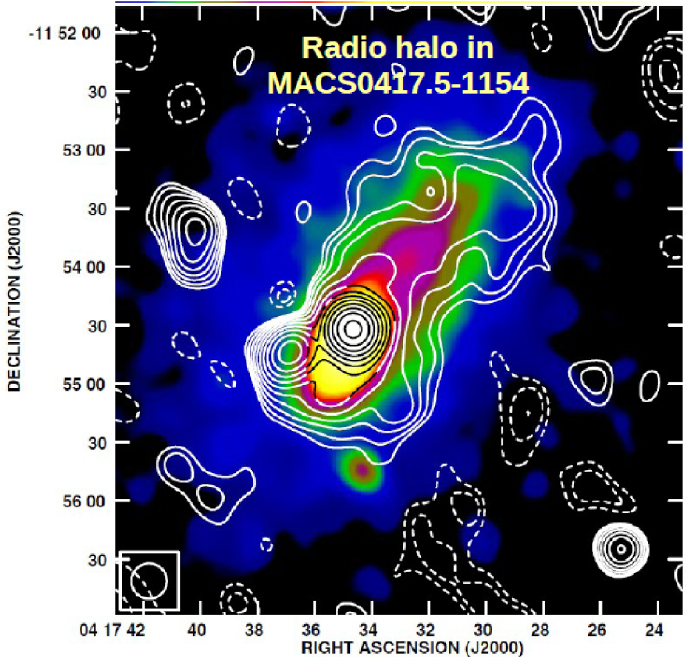}
 \includegraphics[height=6 cm]{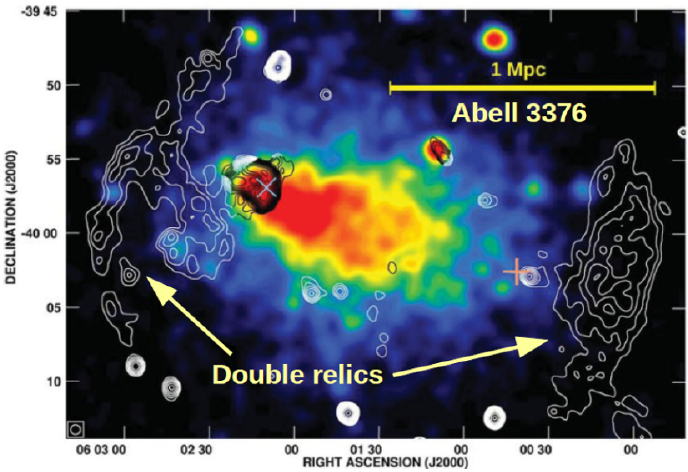}
   \includegraphics[height=6 cm]{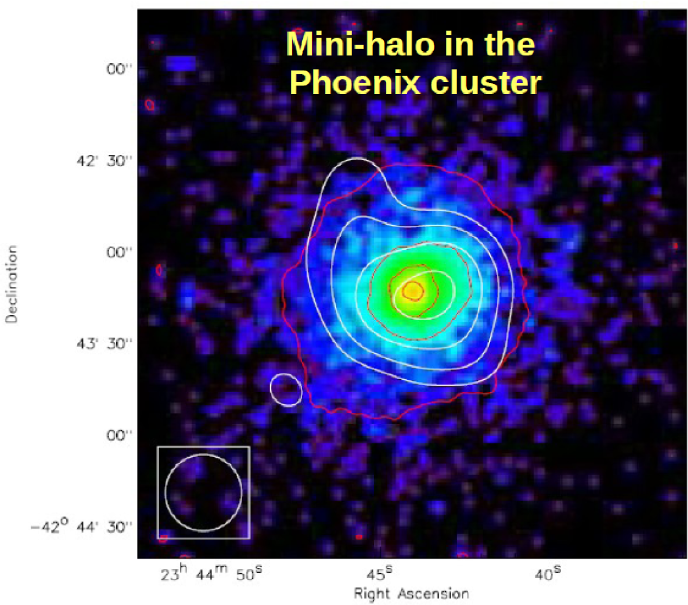}
 \caption{Examples of diffuse radio emission in clusters discussed in this chapter. The X-ray images are 
 shown in colour scale and radio images in white contours are overlaid.
 Top left: Radio halo in the cluster MACS0417.5-1154. Image adapted from \citep{2011JApA...32..529D}. Top right: 
 Double radio relics in the cluster A3376. Image adapted from \citep{kal12}. Bottom: Radio mini-halo in 
 the Phoenix cluster. Red contours show the X-ray emission. Image adapted from \citep{2014ApJ...786L..17V}.
 \label{halorelic}}
\end{figure}

\section {Diffuse synchrotron radio emission from the ICM}
\label{diffuse-synch}
%\subsection{Classification}
 Cluster-wide non-thermal radio emission
has been a topic of study for nearly half a century since its discovery in 
the Coma cluster \citep{1959Natur.183.1663L,1970MNRAS.151....1W}. 
Over the last few decades, a large number of galaxy clusters were imaged at radio
wavelengths ($20 - 200$ cm) using a variety of synthesis radio telescopes like the Very Large
Array (VLA), the Westerbork Synthesis Radio Telescope (WSRT) and the Giant Metrewave
Radio Telescope (GMRT). A large variety in the size, morphology and spectrum of these sources 
has been found. Based on the properties of the radio sources and the host cluster, 
the sources have been classified into three
\footnote{A fourth class of diffuse radio emission are the so-called ``radio phoenixes'' which are smaller scale sources 
that have been proposed to be fading and/or re-accelerated remnants of radio galaxy lobes; these are not discussed in 
this work.} main classes \citep[e.g.][]{fer12}:
\begin{itemize}
 \item {\it Radio halos} -- extended sources of sizes $\sim$ Mpc, nearly co-spatial with X-ray 
 emission from the central regions 
 of clusters.
 \item {\it Radio relics} -- elongated, filamentary or sometimes arc-like sources of few hundred to $1-2$ Mpc on 
 the longer sides, found at the peripheries of clusters. 
 \item {\it Radio mini-halos} -- extended sources of regular morphologies with sizes  $\sim$100 - 500 kpc, surrounding central 
 elliptical galaxies in cool-core clusters.
\end{itemize}
Illustrative examples of these sources are shown in Fig.~\ref{halorelic}.
The common components of the ICM which lead to these sources are the relativistic electrons and magnetic fields. 

The radio spectra of these sources are typically described by a power-law, $S_\nu \propto \nu^{-\alpha}$, 
where $S_\nu$ is the flux density at frequency $\nu$ and $\alpha$ is the spectral index. 
The spectral index, $\alpha$ is related to the injection spectral index, $\delta$ of the electron energy 
distribution, $N(E)dE = N_0 E^{-\delta} dE$ through, $\delta= 2\alpha+1$ (see textbooks such as
\citet{2000thas.book.....P} for details). The spectra originating as a power-laws in the case of 
standard models of particle acceleration can be affected by the effects of overlapping regions of different 
magnetic fields, 
age of the electron population, energy losses by other mechanisms such as the IC losses 
and continuous or intermittent acceleration mechanism. Radio observations over a large range of 
frequencies are required to study the spectral properties of radio halos, relics and mini-halos 
and infer the state of the relativistic electron population and magnetic fields. Below we discuss each 
of these in detail.

The sizes of several hundreds of kiloparsecs in the case of the radio halos imply the role of in-situ 
mechanisms of cosmic ray production 
in the ICM. The primary electron model invokes 
 re-acceleration of electrons from the thermal pool or fossil mildly relativistic electrons via magneto-hydrodynamic (MHD) turbulence injected by 
 cluster mergers \citep{pet01,bru01,pet08,2016MNRAS.tmp..288B}. The secondary or hadronic model invokes 
 collisions between relativistic protons and thermal protons for producing 
 relativistic electrons in the ICM as secondary products along with associated gamma rays 
 \citep[e.g.][]{den80,bla99}. 
 Based on the upper limits on the gamma ray fluxes from clusters, simulations show that the 
 relativistic electrons produced via hadronic models alone are insufficient 
 to explain the radio halos \citep[e.g.][]{don13a} but can contribute to forming what are called `off-state' radio halos 
 in low mass clusters and relaxed clusters \citep{bru11,bro11}.
 
 The radio emission in relics shows another mode of acceleration of mildly relativistic electrons to ultra relativistic energies via acceleration by shocks.
 In the large scale structure there are shocks in the filaments that have Mach numbers of 100s and near cluster peripheries shocks 
 of Mach numbers of a few (1.5 -4) are found \citep[]{min00,2003JKAS...36..111K}. Diffusive shock acceleration (DSA) has been 
 the proposed model for producing relativistic electrons at the shocks  in sources such 
 as the radio relics \citep[e.g.][]{ens98}.
 
 Mini-halos are diffuse sources around the central dominant galaxy in relaxed clusters. 
 The secondary electron model and/or re-acceleration of fossil relativistic electrons by turbulence in the 
 cool core produced due to the infall of sub-clusters inducing sloshing motions 
 have been considered for producing the relativistic electrons responsible for the mini-halos 
 \citep[e.g.][]{git02, pfr04,fuj07,kes10,zuh13}. A connection between cold-fronts and the 
 mini-halos has also been found \citep{maz08,hla13,gia14,2014ApJ...795...73G}. 
 The connection between the central galaxy and the mini-halo and the role of external interaction 
 with the cool-core are open questions being addressed using new radio observations and simulations.

\subsection{Past studies}
\subsubsection{Radio surveys}
The radio observations in the past couple of decades have provided a glimpse of the properties and 
occurrence rates of the diffuse radio sources in galaxy clusters. 
The initial discoveries of the radio halos and relics came mainly from the inspection of 
all sky radio surveys such as the NRAO VLA Sky Survey (NVSS,\citet{con98}). It is a survey at 1.4 GHz 
carried out using the VLA in D configuration (27 antennas in 1 km$^2$ area) 
providing images of the sky with rms sensitivity of 0.45 mJy beam$^{-1}$ where beam$=45''\times45''$. 
A number of radio halos and relics in clusters at redshifts $< 0.2$ were discovered 
using NVSS and subsequently confirmed with deep follow-up observations
 \citep[e.g.][]{gio99,bag02}. 

The low surface brightnesses, $\sim 0.2 - 1\mu$Jy arcsec$^{-2}$ at 1.4 GHz of 
radio halos, relics and mini-halos and their extents of several arcminutes to a degree 
make them a challenge for imaging with radio telescopes. Aperture synthesis radio telescopes 
require a combination of short and long baselines to effectively image the extended sources 
and to separate out the contamination by discrete compact sources. A low frequency aperture synthesis telescope 
 providing the necessary combination of short and long baselines with sufficient sensitivity 
to detect faint structures is ideal to carry out tailored surveys of galaxy clusters. 
 
In the past decade systematic surveys of galaxy clusters at low radio frequencies ($<$GHz) have been taken up 
mainly with the GMRT due to its suitable configuration and sensitivity to extended 
emission. The ``GMRT Radio Halo Survey'' (GRHS) and its extension, the 
``Extended GMRT Radio Halo Survey'' carried out at 610 MHz, are a systematic deep radio survey
of 64 most X-ray luminous ($L_{\mathrm{X}[0.1-2.4\mathrm{keV}]}>5\times10^{44}$ erg s$^{-1}$) galaxy clusters in 
the redshift range 0.2 - 0.4. In the GRHS+EGRHS about sample $22\%$ galaxy clusters 
contain radio halos, $16\%$ 
contain mini-halos and $5\%$ contain radio relics \citep{ven07,ven08,kal13,kal15}. 
Moreover, the upper limits on the non-detections have showed that 
there is a bi-modality in the clusters such that the detections and non-detections are well separated implying 
``radio bright'' and ``radio quiet'' \footnote{This terminology is not connected in any way to the one 
used in the context of radio galaxies.} clusters in the $P_{1.4\mathrm{GHz}}- L_X$ plane 
. Apart from X-ray flux-limited samples of galaxy clusters that have been 
explored so far, with the telescopes
such as the Planck, the South Pole Telescope and the Atacama Cosmology Telescope that detect clusters using the 
Sunyaev- Ze'ldovich effect, mass limited samples can now be explored. 
Radio surveys of complete mass limited samples are ongoing and will provide results in the coming 
years \citep{bas12,cuc15,bon15}.

 \subsubsection{Radio halos and relics}
 The observations of radio halos and relics provide the constraints for the 
 particle acceleration mechanisms. 
 The VLA in C and D configurations, the Westerbork Synthesis Radio Telescope (WSRT) and the 
 GMRT have provided the most sensitive measurements of the diffuse radio emission in 
 galaxy clusters in the last two decades. Spectral index distribution across the extents of radio halos is  
 essential to trace the in-situ acceleration mechanisms and the ageing of the electrons. For this measurement,  
 matched arrays at multiple frequencies are needed and were achieved using a combination of telescopes.
 The first spectral index maps of radio halos were made with the VLA and the WSRT \citep[e.g.][]{gio93,fer04}.
 A spectral index study of the radio halo and relic in the cluster A2256 was carried out with a combination 
 of the GMRT at 150 MHz, the WSRT at 350 MHz and the VLA at 1400 MHz \citep{kal10}. A spectral steepening 
 in the halo at lower frequencies was found in the above observations for the first time and was interpreted in terms of the 
 two epochs of particle acceleration driven by two mergers. Deeper studies with the Karl G. Jansky VLA (JVLA) 
 have further shed light on the spectral properties of the radio relics \citep{owe14,tra15}.
 The follow-up of MACS 
 (Massive Cluster Survey,\citep{ebe10}) clusters with the GMRT have led to the discovery of a halos, relics and mini-halos in couple of 
 clusters and a number of other diffuse sources \citep[][]
 {2011JApA...32..529D, 2013A&A...557A.117P,2015sf2a.conf..247P,2016arXiv160802796P}. 
 The MACS clusters have also been observed at 2 - 24 GHz with the Australia Telescope Compact Array (ATCA) to study the 
 non-thermal and thermal pressures using diffuse emission and the Sunyaev Zel'dovich effect (SZE) \citep{2016Ap&SS.361..255M}.
 
 The SZE has emerged as a promising probe of discovering new clusters 
 in the past few years. The Planck satellite, the South Pole Telescope and the Atacama Cosmology Telescope 
 have all led to discoveries of new clusters \citep{2011A&A...536A...8P,2014A&A...571A..29P,2013ApJ...763..127R,2015ApJ...803...79L}. 
 The first system of radio halo and relic was discovered in a massive new Planck cluster by \citep{bag11} and 
 were followed by discoveries in other clusters \citep[e.g.][]{2013ApJ...766...18G,bon15}. These newly discovered clusters 
 have flatter entropy profiles at the cores and are more likely to be disturbed 
 given the fact that these were missed in the X-ray flux limited catalogues due to their more disturbed morphology. 
 Thus a search for new radio halos and relics in these clusters is on and is furthering the understanding of the 
 role of mass and the dynamical state of the cluster in the generation of radio halos and relics.

 Spectral and polarization studies of relics are essential to distinguish between the models for their origin.
 The prototype of cluster peripheral double relics in the cluster A3376 were discovered and proposed to be 
 the signatures of cluster merger shocks \citep{bag06}. The spectral and polarization study of the A3376 double radio 
 relics led to the observational evidence of its connection to merger shocks \citep{kal12}. The low frequency GMRT
 observations have been crucial in the discoveries of new radio relics 
 \citep{wee10,wee11b,2015MNRAS.453.3483D} and  
 radio halos \citep[e.g.][]{2013ApJ...766...18G,2014ApJ...781L..32V,2015sf2a.conf..247P}.
 The first among the class of ``ultra-steep spectrum'' radio halos predicted by the turbulent re-acceleration model 
 was found in the cluster A520 using the GMRT \citep{bru08}.
 
 \begin{figure}
\centering
   \includegraphics[height=8 cm]{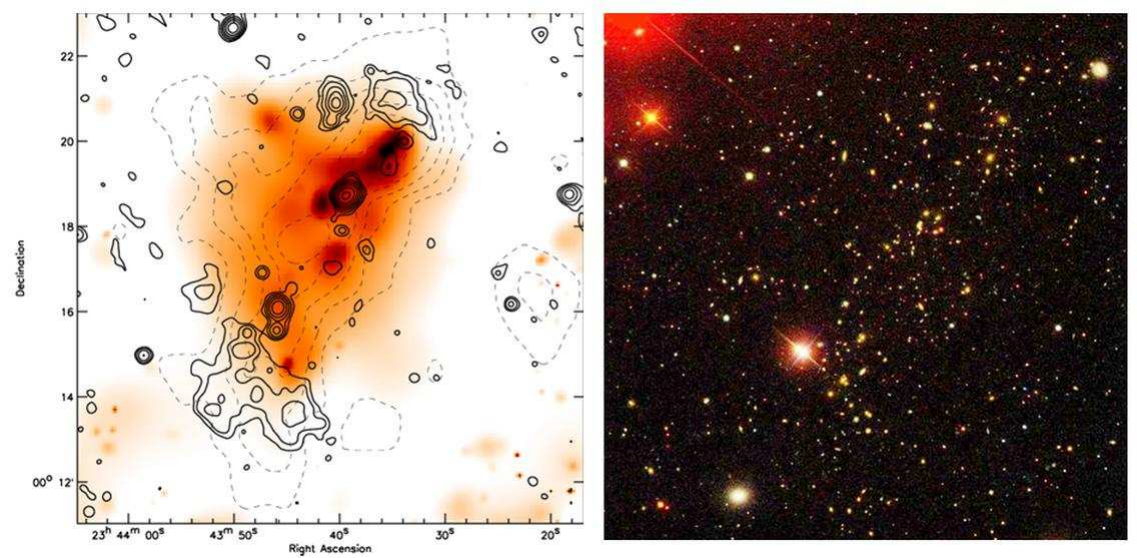}
 \caption{Peculiar radio relics found  in  the filamentary merging cluster
ZWCL 2341.1+0000 \citep{bag02,wee09}.  
Left panel shows the  radio emission  at
610 MHz from  GMRT  (solid contours),   smoothed to 
a circular beam of $15''$  to better show the diffuse radio emission.  
X-ray emission distribution  from  XMM-Newton observations 
is shown in the background.  Dashed contours show the galaxy iso-density
contours from  Sloan Digitized Sky Survey.  The  SDSS colour image of  cluster ZwCl 2341.1+0000
is shown on the  right panel.
 \label{zwcl}}
\end{figure}
 
 \citet{bag02} discovered the diffuse radio emission associated in a complex system of merging
groups of galaxies ZWCl 2341.1+0000 in 1.4 GHz NVSS survey, and concluded that this emission was the first 
evidence of cosmic-ray particle acceleration taking place at magnetized cosmic shocks in a highly filamentary 
environment that is likely in the process of ongoing structure formation.
 In recent years more observations of this unusual and highly filamentary merging 
 structure (Fig.~\ref{zwcl}) and a few other systems have revealed some intriguing details (discussed below) 
 which brought to light unusual 
 aspects of particle acceleration and radio halo and relic formation in elongated, 
 filamentary structures:
\begin{enumerate}[a.]
\item The unusually flat radio spectra of the peripheral double relics ($\alpha < 0.5$) 
implies high Mach number for the shocks, in contradiction with Mach number estimated from
X-ray observations \citep{bag02,wee09,ogr14}.
\item It is suggested that relics and radio halo in 
ZWCl 2341.1+0000 highlight the shortcomings or failure of DSA theory in explaining 
the particle injection spectrum in low mass, less energetic filamentary structures 
of the cosmic-web \citep{ogr14}. Possibly our understanding of the origin of radio relics is 
incomplete, and that non-linear effects are required to explain particle acceleration 
at weak shocks.
\item If radio relics result from particle (re-)acceleration at shock fronts, then this 
shock should span the whole length of the radio relic it traces. Yet in ZWCL 2341.1+000, 
this expectation is not met, where the SE shock front was found to subtend an arc that 
is only about a third of the length of the arc subtended by the radio relic; the result 
was confirmed with a confidence level of $\sim90\%$ by \citep{ogr14}.  Again 
this result is at odds with the standard DSA theory and needs more investigation with deeper 
radio observations and modeling.
 \end{enumerate} 
 
 The GMRT observations at multiple low frequency bands have been used to distinguish between merger shock related 
 relics and adiabatically compressed lobes of radio galaxies.
 The adiabatic compression model proposed by \citet{ens01} was used to model the spectra of the diffuse emission in A754 
 and the relics of double radio galaxies
 \citep{2009ApJ...698L.163D,kal09}. In a multi-wavelength study of radio relics A4038, A1664 and A786, it 
 was found that these are relics not associated with shocks but are adiabatically compressed or fading remnants of radio galaxy lobes \citep{kaldwa12}. The GMRT low frequency observations revealed that the radio relic in A4038 is extended to 
 over 100 kpc and what was seen earlier from observations at frequencies 1.4 GHz was only the brightest portion.

 \subsubsection{Mini-halos}
 Radio mini-halos are diffuse sources, typically of sizes 100 - 500 kpc that surround the 
 central galaxies in cool-core clusters. The central galaxy always has associated compact source 
 at its centre but is not connected to the surrounding diffuse mini-halo with jets but may have a
  role in injecting the seed relativistic electrons \citep[e.g.][]{gia14}. 
 The secondary electrons due to hadronic collisions may play a role in generating the seed electrons that 
 form the mini-halo but are shown to be insufficient to power the mini-halo 
 \citep[][]{zuh14}. It has been proposed that the MHD 
 turbulence in the cool-core, generated via cooling flow or sloshing cores can lead to re-acceleration of relativistic 
 electrons \citep[e.g.][]{git02,maz08}. 
 
 As of now the number count of mini-halos is still limited to about 22 \citep[][Pandey-Pommier et al. 2016, submitted]{gia14}.
 About $50\%$ of the cool-core subsample of the clusters in the Extended GMRT Radio Halo Survey were 
 found to host a mini-halo and a scaling between their radio power and 
 the X-ray luminosity was explored \citep{kal13}. The cooling at the centres and the mini-halos 
 may be connected through scaling relations, but these have not been established due to 
 the lack of statistical samples \citep[e.g.][]{bravi16}. 
 
 Low mass clusters have not been surveyed to search for mini-halos; however there are cases such as the 
 diffuse mini-halo like radio source in the cluster MRC0116+111 which may be remnants of the radio 
 mode activity of the central galaxy \citep[][]{2009MNRAS.399..601B}.

 \subsection{Simulations}
\begin{figure}
\centering
 \includegraphics[width=\textwidth]{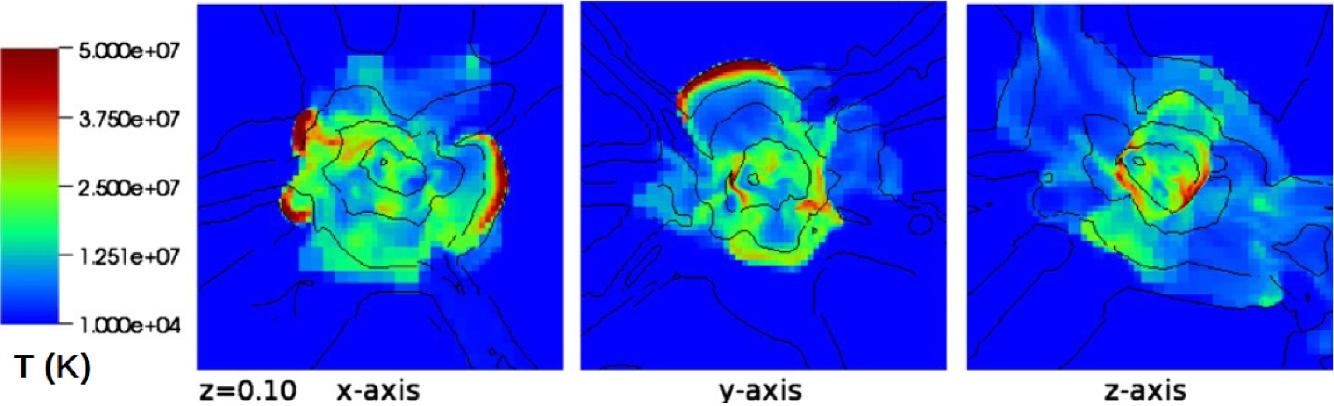}
 \caption{Evolution of the shock in temperature in a simulation by \citet{pau11}, as seen from slices in three different planes of the computational volume. The panels refer to z = 0.1 and to slices
perpendicular to the x-axis (left), to the y-axis (central), and to the z-axis (right), respectively. Each panel has a size of $7.7\times7.7$ Mpc $h^{-1}$ and is cut along the center of mass of the system.\label{paul}}
\end{figure}

Cosmological simulations using 
computations performed with the grid-based, adaptive mesh refinement hydrodynamical code {\it ENZO} show 
that turbulence in clusters can be sustained on timescales exceeding a Gyr \citep{pau11}. 
In the same work it was proposed that the breaking of shocks (seen as high temperature 
structures in the simulation slices) at inflowing filaments around the cluster can produce notches in the radio relics 
(Fig.~\ref{paul}). Simulations 
including magnetic field evolution along with the structure formation and radiative cooling can provide 
predictions for the radio emission from the large scale structure based on the recipes for production of 
relativistic electrons \citep[][]{2007MNRAS.375...77H, don13a,vaz14,2015A&A...580A.119V}. 
It has been shown that a saturation of magnetisation can be achieved in a fully turbulent medium and there 
is near equipartition between 
the magnetic field and kinetic energy densities \citep[e. g.][]{sub06,2012MNRAS.423.2781I}. Magnetic field 
can be evolved in the simulation using the estimate of turbulence and a scaling between turbulent and magnetic energies.
This has been implemented by Paul et al. (2016, in prep.) in the case of Coma and shows agreement with the observed magnetic field profiles \citep{2010A&A...513A..30B} 
based on Faraday rotation of background sources. Radio mini-halos due to electron reacceleration by minor merger driven 
turbulence has been simulated by \citet{zuh13}. Implementation of diffusive shock acceleration and turbulent acceleration 
to accelerate cosmic rays is being used to predict synchrotron emission that will be observable with the SKA.
Predictions by Paul et al. (2016, in prep.) indicate that the estimate of surface brightness at 1.4 GHz with a beam of 
$20''$ is about $10^{-8}$Jy beam$^{-1}$ from the filaments if only turbulent reacceleration is considered
and in the range $10^{-6}$ Jy beam$^{-1}$ if DSA is also included and denser regions such as groups and cluster 
outskirts are considered. The uncertainties such as efficiency of turbulent reacceleration and magnetization in the large scale structure remain and will be constrained by deeper observations that will be made possible by the SKA.
 
\subsection{Prospects for the SKA}
In this section we discuss the numerous possibilities of studying diffuse radio emission from galaxy clusters 
opened by SKA, thanks to its unprecedented sensitivity.
\subsubsection{Deep continuum and polarization imaging of halos, mini-halos and relics}
Radio halos, relics and mini-halos in general have low surface brightness ($\sim1\mu$Jy arcsec$^{-2}$ at 1.4 GHz) 
and among these, radio halos have the lowest brightness levels. The average surface brightness of the known radio halos 
at 1.4 GHz is $\sim0.25\mu$Jy arcsec$^{-2}$ \citep[inferred using the data in][]{fer12}. The
\low will achieve a resolution of $7''\times7''$ at 110 MHz (Table ~\ref{skasens}). 
In order to map the extended sources, 
the images are produced at a lower resolution than the best offered by the telescope configuration. 
Therefore we assume a canonical resolution of $10''\times10''$ for multi-wavelength observations of 
diffuse radio sources with the 
\low and \smid and discuss the predictions.

\begin{figure}
\centering
 \includegraphics[width=15 cm]{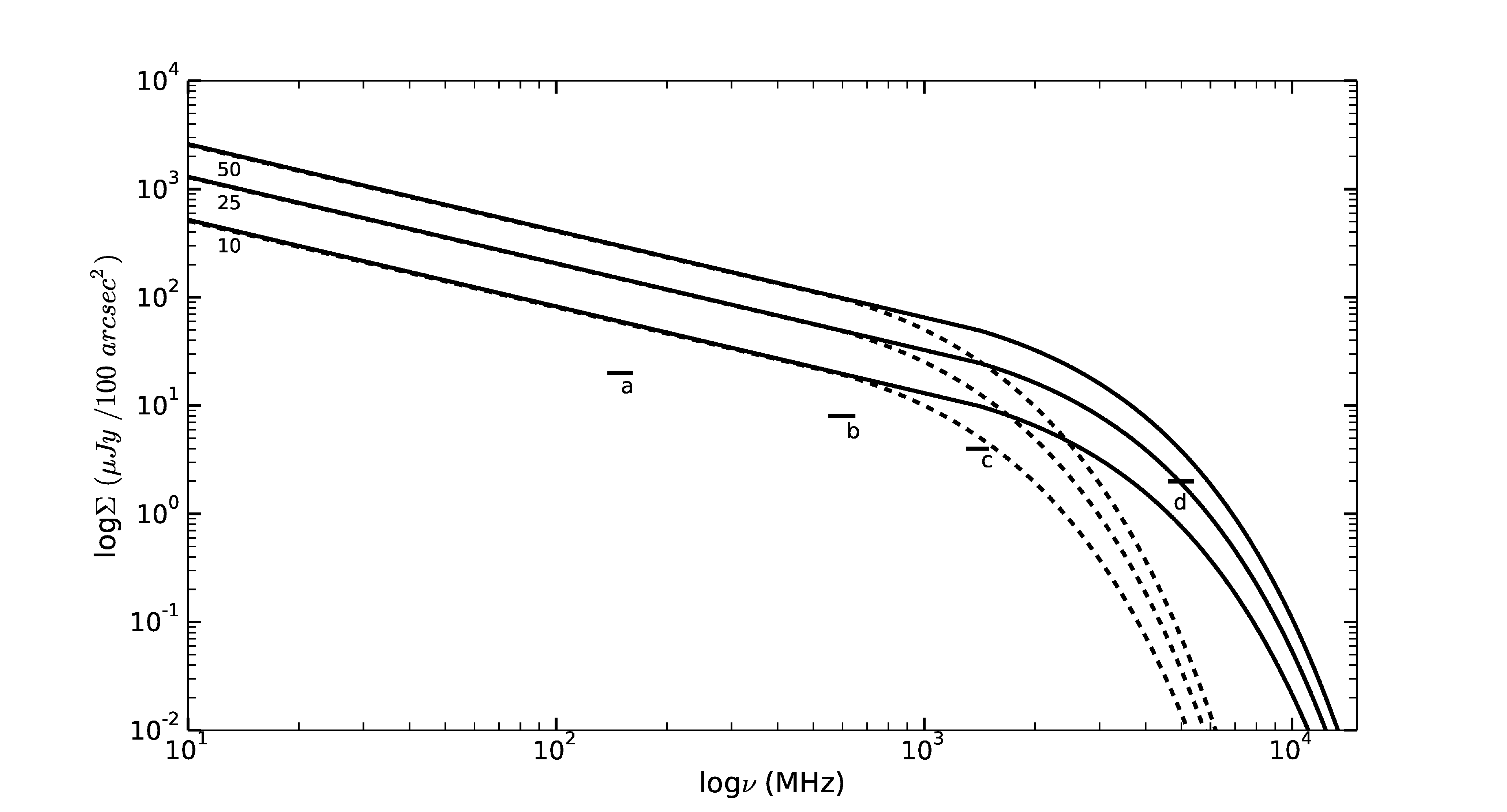}
 \caption{Surface brightness of radio halos as a function of frequency.
 The lines indicate the expected spectra for radio halos that are 
 a combination of a power-law and an exponential decay. The critical frequency 
 where the exponential decay starts is 1.4 GHz (solid lines) and 0.6 GHz (dashed lines), respectively.
The three sets of spectra are for three different 
surface brightness values, 10, 25 and 50 $\mu$Jy / 100 arcsec$^{2}$ for the 
halos at 1.4 GHz. 
The short horizontal lines labelled a, b, c and d indicate the expected confusion limits 
($\mu$Jy / 100 arcsec$^{2}$) at the frequencies $0.15, 0.6, 1.4$ and $5$ GHz, respectively. \label{rhpred}}
\end{figure}

The currently prevalent models of turbulent re-acceleration for radio halos predict that the spectra 
change from a power-law to an exponential decay beyond a certain critical frequency \citep{bru14}. 
Assuming the critical frequency to be 1.4 GHz, the spectrum
corresponding to the mean value of surface brightnesses of known halos
is plotted in Fig.~\ref{rhpred}. The rms confusion at cm wavelengths assuming a Gaussian 
beam is approximated as \citep{2002ASPC..278..155C}, 
\begin{equation}
 \bigg(\frac{\sigma_c}{\mathrm{mJy \,\,beam}^{-1}}\bigg) \approx 0.2 \,\,\bigg(\frac{\nu}{\mathrm{GHz}}\bigg)^{-0.7} \bigg(\frac{\theta_{b}}{\mathrm{arcmin}}\bigg)^2.
\end{equation}
The expected rms confusion at the four frequencies of 0.15, 0.6, 1.4 and 5 GHz (Fig.~\ref{rhpred}) are 
shown in comparison to the radio halo spectra.
It is clear from this figure that most of 
the known halos can be imaged up to 5 GHz if the cut-off frequency is
$\sim$ 1.4 GHz, but, only up to 2 GHz, if the cut-off occurs earlier.
The surface brightnesses of diffuse sources are not uniform. The profiles of radio halos 
show that the central regions are the brightest with a gradual fall towards the edges. 
With \low and \smid, the known radio halos, relics and mini-halos can be imaged to a factor of a few 
deeper. The profiles of the radio halos in the outskirts of the clusters will provide clues to the underlying 
cosmic ray and magnetic field profile with implications to the proposed models for the generation of 
radio halos. It will be possible to reliably map spectral indices over the extent of the radio halos. 
This will lead to constraints on the in-situ acceleration mechanisms at work in the cluster.
The case of radio halos discussed here is an illustration of the weakest of the diffuse radio sources in clusters.
Radio relics and mini-halos being a few times brighter in surface brightness will be imaged in detail 
 leading clues to their origin and evolution.

Furthermore, polarization is a direct probe of the magnetic field 
in the plane of the sky and complements the line of sight magnetic field estimated via Faraday rotation of the 
plane of 
polarization of background sources. 
Through polarization observations coherence lengths of magnetic field can be inferred 
and that is used to refine the simulations. 
Radio halos typically have low polarized emission fraction $<5\%$ or so \citep{fer12}. 
Polarization detection claimed in the radio halo of the cluster A2255 but it is likely due to 
polarized filament in the periphery of the cluster seen projected on the cluster
\citep[][]{2005A&A...430L...5G, 2011A&A...525A.104P}. Radio mini-halos are at cluster centres and thus 
depolarization is expected due to the ICM and hence little is known about their polarization properties.
Radio relics are known to be polarized up to $10-30\%$, as expected based on their origin in shock acceleration 
\citep[see][for reviews]{fer12,bru14}. \smid sensitivities will allow to probe the polarization 
properties of these sources to much deeper 
levels than currently possible and provide important constraints on magnetic fields in clusters of galaxies.

\subsubsection{Constraining high frequency spectra of radio halos, relics and mini-halos}
The spectra of radio halos, relics and mini-halos are important to constrain models proposed to 
explain them. The turbulent re-acceleration model predicts that the spectra will steepen exponentially 
beyond a break frequency decided by the turbulent energy budget in the cluster \citep{cas05,cas06}. 
A handful of sources have been imaged at more than three frequency bands below 1 GHz \citep{fer12} 
and at frequencies higher than a GHz, 
the spectra are largely unknown. Recently two relics were imaged in the frequency range 0.15 - 30 GHz
using a number of radio telescopes and its spectrum shows departure from the expectations 
from the DSA model \citep{2016MNRAS.455.2402S} but theoretical 
works argue that DSA can explain it under certain condition of a shock passing through a cloud of fossil population of 
relativistic electrons \citep{2016arXiv160203278K}. 
It is important to probe the spectra of these diffuse sources at $>$GHz frequencies to 
locate the spectral breaks in them and to test the models. 

Characterisation of galaxy clusters at $>10$ GHz is hampered by the presence of the SZE.
The distortion of the spectrum of the CMB spectrum due to
inverse Compton scattering of the CMB photons by the thermal electrons in the ICM is known as the 
SZE \citep{1972CoASP...4..173S}. At frequencies in the range of tens of GHz, 
the CMB produces a negative signal that can be mixed with the positive radio emission due to radio halos. 
It is possible to model the signal due to SZE as it has a $\nu^{2}$ dependence and study the radio emission. 
Studies with ATCA at frequencies 9 and 18 GHz have shown that radio halos (also relics) such as that in the Bullet cluster 
can be detected, though mixed with the SZE \citep{2010arXiv1005.1394M,2011JApA...32..541M,2016Ap&SS.361..255M}.  

Observations at matched resolution at high and low frequencies are needed to accurately study the radio halos and 
the SZE at GHz frequencies. We discuss the expectations for \smid using the Fig.~\ref{rhpred}.
It can be seen that most of 
the known halos can be imaged up to 5 GHz if the cut-off frequency is
$\sim$ 1.4 GHz, but, only up to 2 GHz, if the cut-off occurs at lower frequencies. 
However the spectra above 2 GHz for the radio halos, relics and mini-halos are largely unknown except in a few cases 
and presents an obvious niche for new observations.

\subsubsection{Search for new radio halos, relics and mini-halos}
The all-sky continuum surveys and targeted surveys of complete samples of clusters 
with the SKA will open windows to discover 100s to 1000s of new radio halos, relics and mini-halos. 
The statistical occurrence of the diffuse radio sources is critical to study their origin and evolution in the context 
of the evolution of the clusters themselves. Current telescopes have allowed a limited number of statistical studies 
owing to the long observations needed to image with high sensitivity \citep{ven07,ven08,kal13,kal15,cuc15,bon15}. 

\begin{figure}
\centering
 \includegraphics[width=10cm,angle=-90]{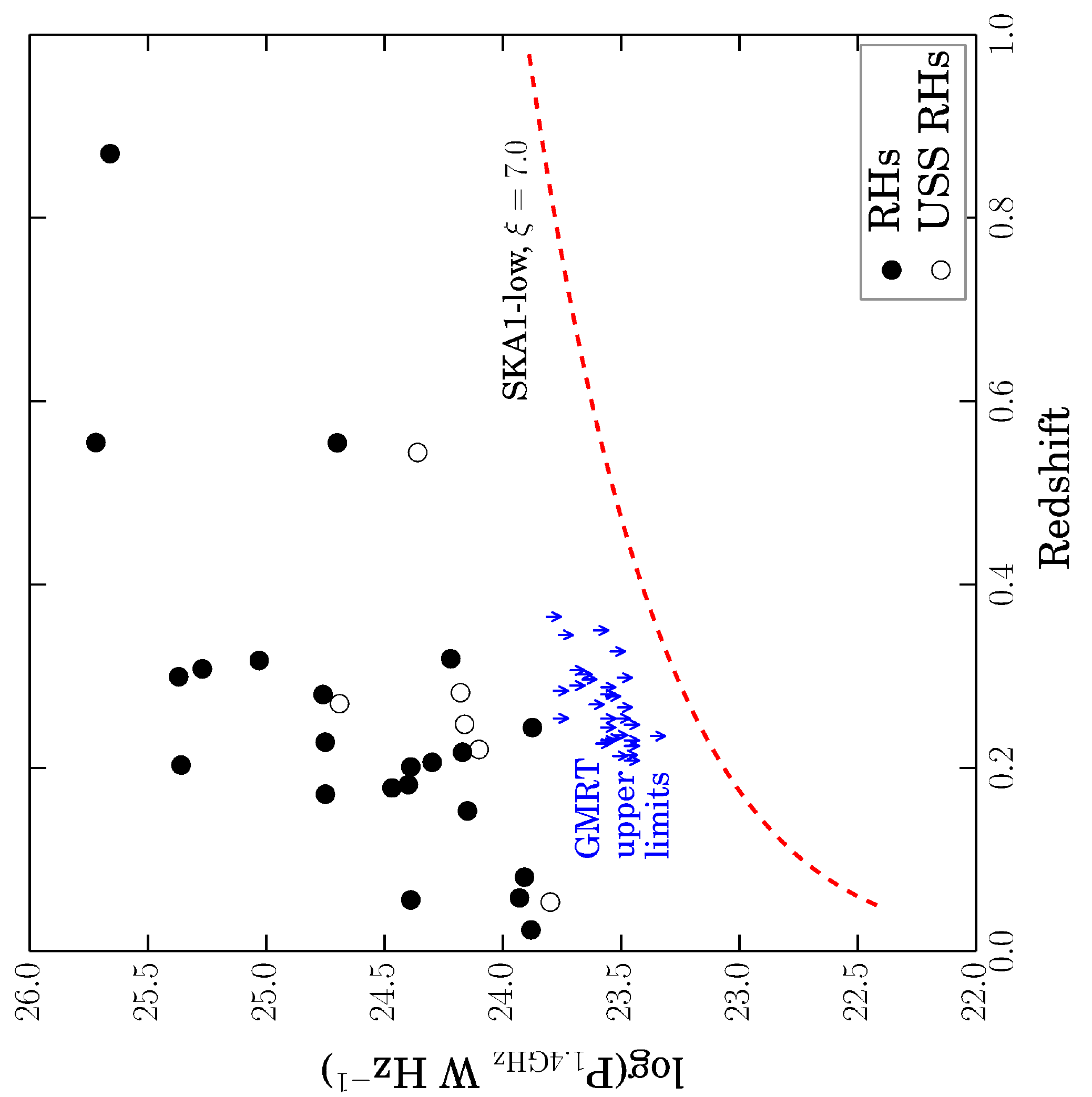}
 \caption{The expected detection thresholds for \low in comparison to the known radio halos. Note that 
 the power at 110 MHz is shown for \low. If it is scaled to power at 1.4 GHz, the expected levels 
 are lower by a factor $(0.11/1.4)^{\alpha}$, where $\alpha$ is the spectral index between 0.11 and 1.4 
 GHz. The known halos from literature and the GMRT upper limits are from \citet{kal15}.\label{rhpred2}}
\end{figure}

The prospects with the SKA to search for such sources in blind surveys and with targeted observations are promising.
From Fig.~\ref{rhpred} it is seen that \low will detect new radio halos as it will 
reach much better sensitivities than the current instruments.
Based on the turbulent acceleration models, it has been predicted that 
about 2600 new radio halos will be discovered in a survey that can reach 
rms 20$\mu$Jy beam$^{-1}$ with a beam $\sim10''$ at 120 MHz \citep{2015aska.confE..73C}. 
From the studies of the known radio halos it has been  inferred that about $58\%$ of the flux density of the radio halo is contained within half its radius 
\citep{bru07}. A minimum detectable flux 
 density of a radio halo, $f_{min}$ can be inferred if it is assumed that a radio halo is detectable when the integrated flux 
 density within half its size gives a signal to noise ratio, $\xi$ \citep{2015aska.confE..73C}. It means that 
 $f_{min}(<0.5\theta_H) \simeq \xi \sqrt{N_b} F_{rms}$, where $N_b$ is the number of independent beams within $0.5\theta_H$.
 It is found that,
 \begin{equation}
  f_{min} \simeq 1.43 \times 10^{-3} \xi \bigg(\frac{\mathrm{F}_\mathrm{rms}}{10\mu\mathrm{Jy}} \bigg) \bigg(\frac{10''}{\theta_b} \bigg) \bigg(\frac{\theta_H(z)}{\mathrm{arcsec}} \bigg).
 \end{equation}
The minimum power of radio halo detectable as a function of redshift for the case of $\xi=7.0$ is shown in Fig.~\ref{rhpred2}. 
 The redshifts beyond 0.4 are essentially unexplored and SKA will provide the first glimpse of the properties of 
 radio halos at those redshifts. The ``off-state'' of clusters which are below the current GMRT upper limits 
 will be detected using the \low. This is critical to understand the cosmic ray content of galaxy clusters.

\section{Superclusters and filaments}
\label{scs}
Binary clusters and systems of multiple interacting clusters can be found in 
superclusters or as independent systems. Typically such systems may be either 
in a stage before they fall into each other or could have undergone the merger 
and are oscillating. In both these cases it is interesting to study the 
ICM in and around these clusters to probe the dynamics of diffuse matter and 
cosmic rays. 

Superclusters such as the Shapley supercluster (SSC) are among the nearest systems available for study.
About 28 clusters were identified within a volume of $2.5\times10^{5}$ $h^{-3}$Mpc$^3$ -- that is an over density 
 of more than a factor of ten with respect to the mean density of Abell clusters at similar galactic latitudes 
 \citep{1989Natur.338..562S}. SSC is also the richest SC in terms of the presence of X-ray emitting clusters 
 \citep{1991MNRAS.248..101R}. Dedicated studies of this region in optical, radio and X-rays
 have revealed several sub-structures within this SSC \citep[e.g.][and references therein]{2000MNRAS.314..594V}. 
 Recently observations of SSC were carried out with the SKA precursor, Murchison Widefield Array (MWA) with the aim of probing diffuse 
 radio emission at cluster and inter-cluster scales (Kale et al. 2016, in prep.). 
 
 The \low opens up the possibility of probing inter-cluster scale radio sources at even better 
 sensitivity than possible with the current MWA. Apart from SSC other binary clusters and superclusters will 
 be of interest to probe the acceleration of cosmic rays and magnetic fields in these regions.
 Targets of interest can be filaments between interacting clusters detected by Planck \citep{2013A&A...550A.134P}.

Apart from the denser structures like the clusters of galaxies and galaxy groups, 
there are filaments in which a large fraction of the galaxies in the Universe reside.
Filaments also contain a large reservoir of inter-galactic medium called the 
warm hot inter-galactic medium (WHIM) with temperatures in the range $10^{5} - 10^{7}$ K 
that has been processed by accretion shocks and is extremely difficult to detect 
in most wavelength bands. Radio bands stand a chance of detecting it if the shocks 
 accelerate electrons. A detection will be able to probe the cosmic 
rays and magnetic fields in the filaments. Flux densities of $\sim 0.12 \mu$Jy at a redshift 
of 0.15 at 150 MHz have been predicted assuming that primary electrons are accelerated 
at cosmological shock waves \citep{2012MNRAS.423.2325A}. A radio detection of filaments has 
been predicted in regions where magnetic field is about $\sim$10 -100 
nG \citep{2015aska.confE..97V}. Recent MHD simulations using $\emph ENZO$ have predicted that 
a non-detection with the SKA can place constraints on the magnetic energy 
in the WHIM to be less than $\sim 1\%$ of the thermal energy \citep{2015A&A...580A.119V}.

Thus although it seems difficult to detect, a strong case exists for detections of filaments
 surrounding massive clusters. Recently filaments of temperatures $\sim10^7$ K were detected in X-rays
 around the massive and merging cluster Abell 2744 \citep{2015Natur.528..105E}.
 If detected in radio, these can potentially constrain 
 the magnetic field in them.

\section {Tailed radio galaxies as tracers of inter-galactic weather}
\label{head-tail}

\subsection{Previous studies}
\begin{figure}
\centering
  \includegraphics[height=6 cm]{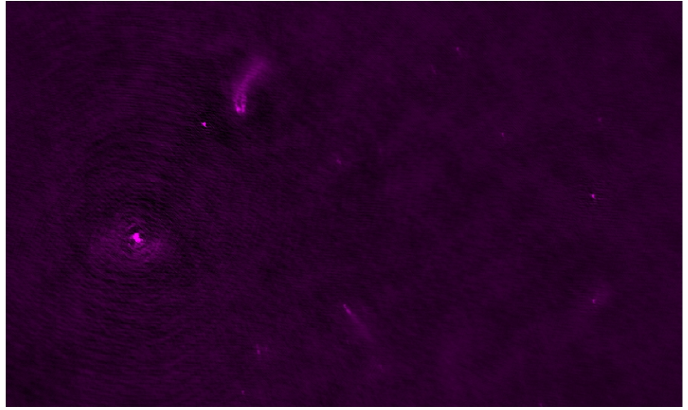}
 \caption{Low frequency, full synthesis GMRT image of head-tail sources in Perseus cluster at 240 MHz. Note the presence of dominant 3C84 radio source and several head-tail radio galaxies, including NGC1265 and IC310.  The field is 4 sq. deg, rms noise is 8.1 mJy beam$^{-1}$ and the angular resolution is 9 arcsec.
 \label{headtail}}
\end{figure}

`Tailed' radio sources \citep{1968MNRAS.138....1R} are
characterised by a head identified with the optical galaxy
and two trails of FR\,I radio source sweeping back from the head 
\citep{1972Natur.237..269M,1973A&A....26..423J}.
These sources are usually found in rich cluster environments,
where jets are understood to have been swept back by the deflecting pressure
of the dense ICM (e.g. Fig.~\ref{headtail}).
Furthermore, the long tails of these galaxies carry the imprint
of relative motion between the non-thermal plasma and the ambient hot gas.
Hence, in the parlance of the field, they reflect the weather conditions of
ICM and the jet dynamics, which allows us to make quantitative statements about
their dynamics and energetics.
Two specific examples of the interaction between radio sources and the ICM are
wide-angle tailed (WAT) and narrow-angle tailed (NAT) radio
sources, where the latter are also addressed as the `head-tail' radio sources.

Recently, a radio study was conducted, using GMRT
for a sample of head-tail radio sources,
concentrating on 3C129, NGC~1265 and IC~310 \citep{2009ASPC..407..157L,2004A&A...420..491L}.  
New X-ray observations were proposed for objects which are in poor environments
where temperature
and abundance variations are likely to be more visible in the X-rays \citep{1994AJ....108.1137R}.
When combined with radio data,
these multi-waveband data probe a variety of studies, namely
(i) collimation and surface brightness of the jets \citep{2014MNRAS.437.3405L,2002MNRAS.336..328L},
(ii) radio jet--ICM interaction \citep{2012ASPC..459..155P,2012IJMPS...8..241P},
(iii) infall of tailed radio sources into the cluster,
(iv) details of cluster merger \citep{2011ApJ...743..199D},
(v) gas pressure to compare with equipartition pressure,
(vi) energy losses, particle acceleration, and
(vii) cluster centre ambiguities.
The potential of such observations is also to reveal
details of cluster mergers such as subsonic $/$ transonic bulk flows,
shocks and turbulence.
Fortunately, the jets survive the encounter
with the ICM, with possible shocks leading to the formation of long
tails, that are devoid of the growth of
Kelvin-Helmholtz instabilities \citep{1995ApJ...445...80L}.

\subsection{Prospects for the SKA}

To illustrate in detail, let us consider two specific science cases related to jets of radio 
galaxies in cluster environment,
(i) is deceleration of jets in these objects is caused by
internal entrainment or if the mass-load of the jet is external
\citep{2012IJMPS...8..241P,2011ApJ...743...42P}
and
(ii) how does the local ICM play a role in shaping
NAT and WAT galaxies 
\citep{1996ApJ...473..651R}.
To answer (i),
observationally the transverse velocity profile of the radio jets
can give clues to understand which of these two processes is more
relevant; if former, no velocity profile across the jet is expected,
whereas in latter case, there would be a velocity profile
with slower layers at the jet boundaries \citep{2002MNRAS.336..328L}.
Next to answer (ii), comparisons between the radio morphologies of
tailed sources in cluster environments and the distribution of
the thermal gas as seen in \textit{Chandra} and \textit{XMM-Newton}
images, indicate that the thermal gas is almost always asymmetric
and aligned towards the direction of the bending \citep{1994ApJ...436...67V}, and many a times
these clusters are undergoing mergers, resulting in large-scale
flows of hot gas owing to the changing gravitational potential \citep{2004MNRAS.351..101K}.
Therefore tailed radio sources in merging clusters 
are diagnostics of the ICM weather and of the evolving
gravitational potential resulting from the merger
as compared to (tailed) radio sources in relaxed clusters.
This argument clearly answers the role played by relaxed and merging clusters
in shaping WATs and NATs.

A series of tiered surveys with increasing sensitivity, but
decreasing in areas are planned to be undertaken with SKA phase-1
(SKA-1). Briefly, the 1--2 GHz band of SKA\,1, with 2~arcsec resolution and
2 $\mu$Jy beam$^{-1}$ will detect several orders of more number of
WAT and NAT radio sources
and hence their parent clusters with whom they are associated.
Clearly, at this resolution and depth we will not only
resolve the radio jet in longitudinal direction, but also in
transverse direction and boost our statistical understanding of at least two
specific science cases discussed above.

\subsection{Building Statistics using SKA}

VLA Fiant Images of the Radio Sky at Twenty-cm (FIRST) survey
listed $\sim$384 WAT and NAT radio sources selected from
3000 deg$^2$ of sky. Of these \citep{2001AJ....121.2915B} have confirmed the existence
of 40 clusters with redshifts up to $z \simeq$ 0.9.
To compare these, we next try
to understand number of WATs and NATs likely to be detected
with SKA surveys.  
The current deep surveys of smaller parts of the sky such as the 
$\sim$4~deg$^2$ of the ATLAS-CDFS (Australia Large Area Survey of
the \textit{Chandra} Deep Field--South) field was performed
using ATCA at 1.4\,GHz \citep{2011PASA...28..215N} with $\sim$11~arcsec resolution
and a depth of 15 $\mu$Jy beam$^{-1}$.  Additionally,
VLA image of the Extended-CDFS, again at 1.4\,GHz with
$\sim$2~arcsec resolution has a depth of 10 $\mu$Jy beam$^{-1}$
\cite{2013ApJS..205...13M}.
Although SKA\,1 survey will be similar to this resolution, it will be 
marginally more sensitive to the VLA--Extended-CDFS
\citep[Figure~5,][]{2014AJ....148...75D},
\citep[Figure~1,][]{2015aska.confE.101J}
and \citep{2015IAUS..313..321J}.  The above surveys detected $\sim$3000 radio sources, of which 
45 are WAT or NAT radio sources and the rest are extended or
(complex) diffuse or ambiguous \citep{2011JApA...32..491D}.  Extrapolating
from these detections, again using \citep[Figure~1,][]{2015aska.confE.101J},
the SKA level surveys will detect
$\sim$10$^6$ WAT and NAT radio sources.
The all-sky radio continuum survey using SKA\,1
will provide more sensitive data to explore a variety of science goals,
including a huge sample, $\sim$10$^6$ clusters of galaxies containing
WAT and NAT radio sources.  Statistical studies of such large samples
of tailed radio sources will provide insights into both the
internal characteristics of astrophysical jets and the
surrounding ICM.

\section{Summary}

This chapter highlights the scientific problems in the area
of galaxy clusters and the cosmic web that can be effectively
addressed using the Square Kilometer Array.
The outstanding problems that need to be addressed are:
\begin{enumerate}[a.]
\item the production and sustenance of relativistic particles and
magnetic fields that are responsible for the observed diffuse
radio emission on cluster-wide scales, 
\item the relative importance of
shocks, turbulent acceleration and hadronic models to the production of 
cluster-wide radio emission, 
\item the detailed properties of the
intra-cluster medium using cluster radio galaxies as tracers, and
\item the faint radio emission arising in structure formation
shocks (the cosmic web).
\end{enumerate}

An underlying theme in the observational study of clusters and
the cosmic web is the detection of low surface brightness and
extended radio emission over a wide range of frequencies like
0.1 - 10 GHz. SKA, with its long baselines and good short spacing
coverage, will be an ideal instrument to detect these features.
It is expected that SKA will reach detection limits that are
a factor of 10 to 50 lower compared to those of the
existing telescopes.  This improved sensitivity is expected to
discover an order of magnitude larger number of clusters
with faint diffuse radio emission hitherto undetected. These 
observations from SKA are expected to revolutionise the field
of clusters and the cosmic web.

\section*{Acknowledgments}
We thank the anonymous referee for providing comments on our paper.
RK and SP acknowledge the support through the DST-INSPIRE Faculty Award.
JB  and  MP gratefully acknowledge generous support from the Indo-French Center
for the Promotion of Advanced Research (Centre Franco-Indien pour la Promotion de la Recherche Avan\'{c}ee) 
under programme no. 5204-2. Research in Astronomy at IIT Indore was made possible through a grant by the institute.
We thank the staff of GMRT, who made these observations possible. The GMRT is run by the National Centre for Radio 
Astrophysics of the Tata Institute of Fundamental Research. 
%%\begin{thebibliography}{}
\bibliographystyle{mn2e}
\bibliography{ruta_all_1,apj-jour,ska,prd,ssc,Sec6.dvl}

\begin{thebibliography}{131}
\expandafter\ifx\csname natexlab\endcsname\relax\def\natexlab#1{#1}\fi

\bibitem[{{Ackermann} {et~al}\mbox{.}(2014){Ackermann}, {Ajello}, {Albert},
  {Allafort}, {Atwood}, {Baldini}, {Ballet}, {Barbiellini}, {Bastieri},
  {Bechtol}, {Bellazzini}, {Bloom}, {Bonamente}, {Bottacini}, {Brandt},
  {Bregeon}, {Brigida}, {Bruel}, {Buehler}, {Buson}, {Caliandro}, {Cameron},
  {Caraveo}, {Cavazzuti}, {Chaves}, {Chiang}, {Chiaro}, {Ciprini}, {Claus},
  {Cohen-Tanugi}, {Conrad}, {D'Ammando}, {de Angelis}, {de Palma}, {Dermer},
  {Digel}, {Drell}, {Drlica-Wagner}, {Favuzzi}, {Franckowiak}, {Funk}, {Fusco},
  {Gargano}, {Gasparrini}, {Germani}, {Giglietto}, {Giordano}, {Giroletti},
  {Godfrey}, {Gomez-Vargas}, {Grenier}, {Guiriec}, {Gustafsson}, {Hadasch},
  {Hayashida}, {Hewitt}, {Hughes}, {Jeltema}, {J{\'o}hannesson}, {Johnson},
  {Kamae}, {Kataoka}, {Kn{\"o}dlseder}, {Kuss}, {Lande}, {Larsson},
  {Latronico}, {Llena Garde}, {Longo}, {Loparco}, {Lovellette}, {Lubrano},
  {Mayer}, {Mazziotta}, {McEnery}, {Michelson}, {Mitthumsiri}, {Mizuno},
  {Monzani}, {Morselli}, {Moskalenko}, {Murgia}, {Nemmen}, {Nuss}, {Ohsugi},
  {Orienti}, {Orlando}, {Ormes}, {Perkins}, {Pesce-Rollins}, {Piron}, {Pivato},
  {Rain{\`o}}, {Rando}, {Razzano}, {Razzaque}, {Reimer}, {Reimer}, {Ruan},
  {S{\'a}nchez-Conde}, {Schulz}, {Sgr{\`o}}, {Siskind}, {Spandre}, {Spinelli},
  {Storm}, {Strong}, {Suson}, {Takahashi}, {Thayer}, {Thayer}, {Thompson},
  {Tibaldo}, {Tinivella}, {Torres}, {Troja}, {Uchiyama}, {Usher},
  {Vandenbroucke}, {Vianello}, {Vitale}, {Winer}, {Wood}, {Zimmer}, {Fermi-LAT
  Collaboration}, {Pinzke}, \& {Pfrommer}}]{2014ApJ...787...18A}
{Ackermann} M. {et~al.}, 2014, \apj, 787, 18

\bibitem[{{Ackermann} {et~al}\mbox{.}(2010){Ackermann}, {Ajello}, {Allafort},
  {Baldini}, {Ballet}, {Barbiellini}, {Bastieri}, {Bechtol}, {Bellazzini},
  {Blandford}, {Blasi}, {Bloom}, {Bonamente}, {Borgland}, {Bouvier}, {Brandt},
  {Bregeon}, {Brigida}, {Bruel}, {Buehler}, {Buson}, {Caliandro}, {Cameron},
  {Caraveo}, {Carrigan}, {Casandjian}, {Cavazzuti}, {Cecchi}, {{\c C}elik},
  {Charles}, {Chekhtman}, {Cheung}, {Chiang}, {Ciprini}, {Claus},
  {Cohen-Tanugi}, {Colafrancesco}, {Cominsky}, {Conrad}, {Dermer}, {de Palma},
  {Silva}, {Drell}, {Dubois}, {Dumora}, {Edmonds}, {Farnier}, {Favuzzi},
  {Frailis}, {Fukazawa}, {Funk}, {Fusco}, {Gargano}, {Gasparrini}, {Gehrels},
  {Germani}, {Giglietto}, {Giordano}, {Giroletti}, {Glanzman}, {Godfrey},
  {Grenier}, {Grondin}, {Guiriec}, {Hadasch}, {Harding}, {Hayashida}, {Hays},
  {Horan}, {Hughes}, {Jeltema}, {J{\'o}hannesson}, {Johnson}, {Johnson},
  {Johnson}, {Kamae}, {Katagiri}, {Kataoka}, {Kerr}, {Kn{\"o}dlseder}, {Kuss},
  {Lande}, {Latronico}, {Lee}, {Lemoine-Goumard}, {Longo}, {Loparco}, {Lott},
  {Lovellette}, {Lubrano}, {Madejski}, {Makeev}, {Mazziotta}, {Michelson},
  {Mitthumsiri}, {Mizuno}, {Moiseev}, {Monte}, {Monzani}, {Morselli},
  {Moskalenko}, {Murgia}, {Naumann-Godo}, {Nolan}, {Norris}, {Nuss}, {Ohsugi},
  {Omodei}, {Orlando}, {Ormes}, {Ozaki}, {Paneque}, {Panetta}, {Pepe},
  {Pesce-Rollins}, {Petrosian}, {Pfrommer}, {Piron}, {Porter}, {Profumo},
  {Rain{\`o}}, {Rando}, {Razzano}, {Reimer}, {Reimer}, {Reposeur}, {Ripken},
  {Ritz}, {Rodriguez}, {Romani}, {Roth}, {Sadrozinski}, {Sander}, {Saz
  Parkinson}, {Scargle}, {Sgr{\`o}}, {Siskind}, {Smith}, {Spandre}, {Spinelli},
  {Starck}, {Stawarz}, {Strickman}, {Strong}, {Suson}, {Tajima}, {Takahashi},
  {Takahashi}, {Tanaka}, {Thayer}, {Thayer}, {Tibaldo}, {Tibolla}, {Torres},
  {Tosti}, {Tramacere}, {Uchiyama}, {Usher}, {Vandenbroucke}, {Vasileiou},
  {Vilchez}, {Vitale}, {Waite}, {Wang}, {Winer}, {Wood}, {Yang}, {Ylinen}, \&
  {Ziegler}}]{ack10}
{Ackermann} M. {et~al.}, 2010, \apjl, 717, L71

\bibitem[{{Araya-Melo} {et~al}\mbox{.}(2012){Araya-Melo}, {Arag{\'o}n-Calvo},
  {Br{\"u}ggen}, \& {Hoeft}}]{2012MNRAS.423.2325A}
{Araya-Melo} P.~A., {Arag{\'o}n-Calvo} M.~A., {Br{\"u}ggen} M., {Hoeft} M.,
  2012, \mnras, 423, 2325

\bibitem[{{Bagchi} {et~al}\mbox{.}(2006){Bagchi}, {Durret}, {Neto}, \&
  {Paul}}]{bag06}
{Bagchi} J., {Durret} F., {Neto} G.~B.~L., {Paul} S., 2006, Science, 314, 791

\bibitem[{{Bagchi} {et~al}\mbox{.}(2002){Bagchi}, {En{\ss}lin}, {Miniati},
  {Stalin}, {Singh}, {Raychaudhury}, \& {Humeshkar}}]{bag02}
{Bagchi} J., {En{\ss}lin} T.~A., {Miniati} F., {Stalin} C.~S., {Singh} M.,
  {Raychaudhury} S., {Humeshkar} N.~B., 2002, ArXiv e-prints, 7, 249

\bibitem[{{Bagchi} {et~al}\mbox{.}(2009){Bagchi}, {Jacob}, {Gopal-Krishna},
  {Werner}, {Wadnerkar}, {Belapure}, \& {Kumbharkhane}}]{2009MNRAS.399..601B}
{Bagchi} J., {Jacob} J., {Gopal-Krishna}, {Werner} N., {Wadnerkar} N.,
  {Belapure} J., {Kumbharkhane} A.~C., 2009, \mnras, 399, 601

\bibitem[{{Bagchi} {et~al}\mbox{.}(2011){Bagchi}, {Sirothia}, {Werner},
  {Pandge}, {Kantharia}, {Ishwara-Chandra}, {Gopal-Krishna}, {Paul}, \&
  {Joshi}}]{bag11}
{Bagchi} J. {et~al.}, 2011, ApJ, 736, L8

\bibitem[{{Basu}(2012)}]{bas12}
{Basu} K., 2012, MNRAS, 421, L112

\bibitem[{{Blanton} {et~al}\mbox{.}(2001){Blanton}, {Gregg}, {Helfand},
  {Becker}, \& {Leighly}}]{2001AJ....121.2915B}
{Blanton} E.~L., {Gregg} M.~D., {Helfand} D.~J., {Becker} R.~H., {Leighly}
  K.~M., 2001, \aj, 121, 2915

\bibitem[{{Blasi} \& {Colafrancesco}(1999)}]{bla99}
{Blasi} P., {Colafrancesco} S., 1999, Astroparticle Physics, 12, 169

\bibitem[{{Bonafede} {et~al}\mbox{.}(2010){Bonafede}, {Feretti}, {Murgia},
  {Govoni}, {Giovannini}, {Dallacasa}, {Dolag}, \&
  {Taylor}}]{2010A&A...513A..30B}
{Bonafede} A., {Feretti} L., {Murgia} M., {Govoni} F., {Giovannini} G.,
  {Dallacasa} D., {Dolag} K., {Taylor} G.~B., 2010, \aap, 513, A30

\bibitem[{{Bonafede} {et~al}\mbox{.}(2015){Bonafede}, {Intema}, {Br{\"u}ggen},
  {Vazza}, {Basu}, {Sommer}, {Ebeling}, {de Gasperin}, {R{\"o}ttgering}, {van
  Weeren}, \& {Cassano}}]{bon15}
{Bonafede} A. {et~al.}, 2015, \mnras, 454, 3391

\bibitem[{{Bond}, {Kofman} \& {Pogosyan}(1996){Bond}, {Kofman}, \&
  {Pogosyan}}]{1996Natur.380..603B}
{Bond} J.~R., {Kofman} L., {Pogosyan} D., 1996, \nat, 380, 603

\bibitem[{{Bravi}, {Gitti} \& {Brunetti}(2016){Bravi}, {Gitti}, \&
  {Brunetti}}]{bravi16}
{Bravi} L., {Gitti} M., {Brunetti} G., 2016, \mnras, 455, L41

\bibitem[{{Brown} {et~al}\mbox{.}(2011){Brown}, {Emerick}, {Rudnick}, \&
  {Brunetti}}]{bro11}
{Brown} S., {Emerick} A., {Rudnick} L., {Brunetti} G., 2011, \apjl, 740, L28

\bibitem[{{Brunetti}(2011)}]{bru11}
{Brunetti} G., 2011, MMSAI, 82, 515

\bibitem[{{Brunetti} {et~al}\mbox{.}(2012){Brunetti}, {Blasi}, {Reimer},
  {Rudnick}, {Bonafede}, \& {Brown}}]{bru12}
{Brunetti} G., {Blasi} P., {Reimer} O., {Rudnick} L., {Bonafede} A., {Brown}
  S., 2012, \mnras, 426, 956

\bibitem[{{Brunetti} {et~al}\mbox{.}(2008){Brunetti}, {Giacintucci}, {Cassano},
  {Lane}, {Dallacasa}, {Venturi}, {Kassim}, {Setti}, {Cotton}, \&
  {Markevitch}}]{bru08}
{Brunetti} G. {et~al.}, 2008, Nature, 455, 944

\bibitem[{{Brunetti} \& {Jones}(2014)}]{bru14}
{Brunetti} G., {Jones} T.~W., 2014, International Journal of Modern Physics D,
  23, 30007

\bibitem[{{Brunetti} \& {Lazarian}(2016)}]{2016MNRAS.tmp..288B}
{Brunetti} G., {Lazarian} A., 2016, \mnras

\bibitem[{{Brunetti} {et~al}\mbox{.}(2001){Brunetti}, {Setti}, {Feretti}, \&
  {Giovannini}}]{bru01}
{Brunetti} G., {Setti} G., {Feretti} L., {Giovannini} G., 2001, MNRAS, 320, 365

\bibitem[{{Brunetti} {et~al}\mbox{.}(2007){Brunetti}, {Venturi}, {Dallacasa},
  {Cassano}, {Dolag}, {Giacintucci}, \& {Setti}}]{bru07}
{Brunetti} G., {Venturi} T., {Dallacasa} D., {Cassano} R., {Dolag} K.,
  {Giacintucci} S., {Setti} G., 2007, ApJ, 670, L5

\bibitem[{{Cassano} {et~al}\mbox{.}(2015){Cassano}, {Bernardi}, {Brunetti},
  {Br{\"u}ggen}, {Clarke}, {Dallacasa}, {Dolag}, {Ettori}, {Giacintucci},
  {Giocoli}, {Gitti}, {Johnston-Hollitt}, {Kale}, {Markevich}, {Norris},
  {Pommier}, {Pratt}, {Rottgering}, \& {Venturi}}]{2015aska.confE..73C}
{Cassano} R. {et~al.}, 2015, Advancing Astrophysics with the Square Kilometre
  Array (AASKA14), 73

\bibitem[{{Cassano} \& {Brunetti}(2005)}]{cas05}
{Cassano} R., {Brunetti} G., 2005, \mnras, 357, 1313

\bibitem[{{Cassano}, {Brunetti} \& {Setti}(2006){Cassano}, {Brunetti}, \&
  {Setti}}]{cas06}
{Cassano} R., {Brunetti} G., {Setti} G., 2006, MNRAS, 369, 1577

\bibitem[{{Condon}(2002)}]{2002ASPC..278..155C}
{Condon} J.~J., 2002, in Astronomical Society of the Pacific Conference Series,
  Vol. 278, Single-Dish Radio Astronomy: Techniques and Applications,
  {Stanimirovic} S., {Altschuler} D., {Goldsmith} P., {Salter} C., eds., pp.
  155--171

\bibitem[{{Condon} {et~al}\mbox{.}(1998){Condon}, {Cotton}, {Greisen}, {Yin},
  {Perley}, {Taylor}, \& {Broderick}}]{con98}
{Condon} J.~J., {Cotton} W.~D., {Greisen} E.~W., {Yin} Q.~F., {Perley} R.~A.,
  {Taylor} G.~B., {Broderick} J.~J., 1998, AJ, 115, 1693

\bibitem[{{Cuciti} {et~al}\mbox{.}(2015){Cuciti}, {Cassano}, {Brunetti},
  {Dallacasa}, {Kale}, {Ettori}, \& {Venturi}}]{cuc15}
{Cuciti} V., {Cassano} R., {Brunetti} G., {Dallacasa} D., {Kale} R., {Ettori}
  S., {Venturi} T., 2015, \aap, 580, A97

\bibitem[{{de Gasperin} {et~al}\mbox{.}(2015){de Gasperin}, {Intema}, {van
  Weeren}, {Dawson}, {Golovich}, {Wittman}, {Bonafede}, \&
  {Br{\"u}ggen}}]{2015MNRAS.453.3483D}
{de Gasperin} F., {Intema} H.~T., {van Weeren} R.~J., {Dawson} W.~A.,
  {Golovich} N., {Wittman} D., {Bonafede} A., {Br{\"u}ggen} M., 2015, \mnras,
  453, 3483

\bibitem[{{Dehghan} {et~al}\mbox{.}(2014){Dehghan}, {Johnston-Hollitt},
  {Franzen}, {Norris}, \& {Miller}}]{2014AJ....148...75D}
{Dehghan} S., {Johnston-Hollitt} M., {Franzen} T.~M.~O., {Norris} R.~P.,
  {Miller} N.~A., 2014, \aj, 148, 75

\bibitem[{{Dehghan} {et~al}\mbox{.}(2011){Dehghan}, {Johnston-Hollitt}, {Mao},
  {Norris}, {Miller}, \& {Huynh}}]{2011JApA...32..491D}
{Dehghan} S., {Johnston-Hollitt} M., {Mao} M., {Norris} R.~P., {Miller} N.~A.,
  {Huynh} M., 2011, Journal of Astrophysics and Astronomy, 32, 491

\bibitem[{{Dennison}(1980)}]{den80}
{Dennison} B., 1980, \apjl, 239, L93

\bibitem[{{Donnert} {et~al}\mbox{.}(2013){Donnert}, {Dolag}, {Brunetti}, \&
  {Cassano}}]{don13a}
{Donnert} J., {Dolag} K., {Brunetti} G., {Cassano} R., 2013, MNRAS, 429, 3564

\bibitem[{{Douglass} {et~al}\mbox{.}(2011){Douglass}, {Blanton}, {Clarke},
  {Randall}, \& {Wing}}]{2011ApJ...743..199D}
{Douglass} E.~M., {Blanton} E.~L., {Clarke} T.~E., {Randall} S.~W., {Wing}
  J.~D., 2011, \apj, 743, 199

\bibitem[{{Dwarakanath} \& {Kale}(2009)}]{2009ApJ...698L.163D}
{Dwarakanath} K.~S., {Kale} R., 2009, \apjl, 698, L163

\bibitem[{{Dwarakanath}, {Malu} \& {Kale}(2011){Dwarakanath}, {Malu}, \&
  {Kale}}]{2011JApA...32..529D}
{Dwarakanath} K.~S., {Malu} S., {Kale} R., 2011, Journal of Astrophysics and
  Astronomy, 32, 529

\bibitem[{{Ebeling} {et~al}\mbox{.}(2010){Ebeling}, {Edge}, {Mantz}, {Barrett},
  {Henry}, {Ma}, \& {van Speybroeck}}]{ebe10}
{Ebeling} H., {Edge} A.~C., {Mantz} A., {Barrett} E., {Henry} J.~P., {Ma}
  C.~J., {van Speybroeck} L., 2010, \mnras, 407, 83

\bibitem[{{Eckert} {et~al}\mbox{.}(2015){Eckert}, {Jauzac}, {Shan}, {Kneib},
  {Erben}, {Israel}, {Jullo}, {Klein}, {Massey}, {Richard}, \&
  {Tchernin}}]{2015Natur.528..105E}
{Eckert} D. {et~al.}, 2015, \nat, 528, 105

\bibitem[{{En{\ss}lin} {et~al}\mbox{.}(1998){En{\ss}lin}, {Biermann}, {Klein},
  \& {Kohle}}]{ens98}
{En{\ss}lin} T.~A., {Biermann} P.~L., {Klein} U., {Kohle} S., 1998, A\&A, 332,
  395

\bibitem[{{En{\ss}lin} \& {Gopal-Krishna}(2001)}]{ens01}
{En{\ss}lin} T.~A., {Gopal-Krishna}, 2001, A\&A, 366, 26

\bibitem[{{Felten} {et~al}\mbox{.}(1966){Felten}, {Gould}, {Stein}, \&
  {Woolf}}]{1966ApJ...146..955F}
{Felten} J.~E., {Gould} R.~J., {Stein} W.~A., {Woolf} N.~J., 1966, \apj, 146,
  955

\bibitem[{{Feretti} {et~al}\mbox{.}(2012){Feretti}, {Giovannini}, {Govoni}, \&
  {Murgia}}]{fer12}
{Feretti} L., {Giovannini} G., {Govoni} F., {Murgia} M., 2012, AApR, 20, 54

\bibitem[{{Feretti} {et~al}\mbox{.}(2004){Feretti}, {Orr{\`u}}, {Brunetti},
  {Giovannini}, {Kassim}, \& {Setti}}]{fer04}
{Feretti} L., {Orr{\`u}} E., {Brunetti} G., {Giovannini} G., {Kassim} N.,
  {Setti} G., 2004, \aap, 423, 111

\bibitem[{{Fujita} {et~al}\mbox{.}(2007){Fujita}, {Kohri}, {Yamazaki}, \&
  {Kino}}]{fuj07}
{Fujita} Y., {Kohri} K., {Yamazaki} R., {Kino} M., 2007, \apjl, 663, L61

\bibitem[{{Giacintucci} {et~al}\mbox{.}(2013){Giacintucci}, {Kale}, {Wik},
  {Venturi}, \& {Markevitch}}]{2013ApJ...766...18G}
{Giacintucci} S., {Kale} R., {Wik} D.~R., {Venturi} T., {Markevitch} M., 2013,
  \apj, 766, 18

\bibitem[{{Giacintucci} {et~al}\mbox{.}(2014{\natexlab{a}}){Giacintucci},
  {Markevitch}, {Brunetti}, {ZuHone}, {Venturi}, {Mazzotta}, \&
  {Bourdin}}]{2014ApJ...795...73G}
{Giacintucci} S., {Markevitch} M., {Brunetti} G., {ZuHone} J.~A., {Venturi} T.,
  {Mazzotta} P., {Bourdin} H., 2014{\natexlab{a}}, \apj, 795, 73

\bibitem[{{Giacintucci} {et~al}\mbox{.}(2014{\natexlab{b}}){Giacintucci},
  {Markevitch}, {Venturi}, {Clarke}, {Cassano}, \& {Mazzotta}}]{gia14}
{Giacintucci} S., {Markevitch} M., {Venturi} T., {Clarke} T.~E., {Cassano} R.,
  {Mazzotta} P., 2014{\natexlab{b}}, ApJ, 781, 9

\bibitem[{{Giovannini} {et~al}\mbox{.}(1993){Giovannini}, {Feretti}, {Venturi},
  {Kim}, \& {Kronberg}}]{gio93}
{Giovannini} G., {Feretti} L., {Venturi} T., {Kim} K.-T., {Kronberg} P.~P.,
  1993, \apj, 406, 399

\bibitem[{{Giovannini}, {Tordi} \& {Feretti}(1999){Giovannini}, {Tordi}, \&
  {Feretti}}]{gio99}
{Giovannini} G., {Tordi} M., {Feretti} L., 1999, New Astr. Rev., 4, 141

\bibitem[{{Girardi} {et~al}\mbox{.}(1993){Girardi}, {Biviano}, {Giuricin},
  {Mardirossian}, \& {Mezzetti}}]{1993ApJ...404...38G}
{Girardi} M., {Biviano} A., {Giuricin} G., {Mardirossian} F., {Mezzetti} M.,
  1993, \apj, 404, 38

\bibitem[{{Gitti}, {Brunetti} \& {Setti}(2002){Gitti}, {Brunetti}, \&
  {Setti}}]{git02}
{Gitti} M., {Brunetti} G., {Setti} G., 2002, \aap, 386, 456

\bibitem[{{Govoni} {et~al}\mbox{.}(2005){Govoni}, {Murgia}, {Feretti},
  {Giovannini}, {Dallacasa}, \& {Taylor}}]{2005A&A...430L...5G}
{Govoni} F., {Murgia} M., {Feretti} L., {Giovannini} G., {Dallacasa} D.,
  {Taylor} G.~B., 2005, \aap, 430, L5

\bibitem[{{Hlavacek-Larrondo} {et~al}\mbox{.}(2013){Hlavacek-Larrondo},
  {Allen}, {Taylor}, {Fabian}, {Canning}, {Werner}, {Sanders}, {Grimes},
  {Ehlert}, \& {von der Linden}}]{hla13}
{Hlavacek-Larrondo} J. {et~al.}, 2013, ArXiv e-prints

\bibitem[{{Hoeft} \& {Br{\"u}ggen}(2007)}]{2007MNRAS.375...77H}
{Hoeft} M., {Br{\"u}ggen} M., 2007, \mnras, 375, 77

\bibitem[{{Iapichino} \& {Br{\"u}ggen}(2012)}]{2012MNRAS.423.2781I}
{Iapichino} L., {Br{\"u}ggen} M., 2012, \mnras, 423, 2781

\bibitem[{{Jaffe} \& {Perola}(1973)}]{1973A&A....26..423J}
{Jaffe} W.~J., {Perola} G.~C., 1973, \aap, 26, 423

\bibitem[{{Johnston-Hollitt}, {Dehghan} \&
  {Pratley}(2015{\natexlab{a}}){Johnston-Hollitt}, {Dehghan}, \&
  {Pratley}}]{2015aska.confE.101J}
{Johnston-Hollitt} M., {Dehghan} S., {Pratley} L., 2015{\natexlab{a}},
  Advancing Astrophysics with the Square Kilometre Array (AASKA14), 101

\bibitem[{{Johnston-Hollitt}, {Dehghan} \&
  {Pratley}(2015{\natexlab{b}}){Johnston-Hollitt}, {Dehghan}, \&
  {Pratley}}]{2015IAUS..313..321J}
{Johnston-Hollitt} M., {Dehghan} S., {Pratley} L., 2015{\natexlab{b}}, in IAU
  Symposium, Vol. 313, Extragalactic Jets from Every Angle, {Massaro} F.,
  {Cheung} C.~C., {Lopez} E., {Siemiginowska} A., eds., pp. 321--326

\bibitem[{{Kale} \& {Dwarakanath}(2009)}]{kal09}
{Kale} R., {Dwarakanath} K.~S., 2009, \apj, 699, 1883

\bibitem[{{Kale} \& {Dwarakanath}(2010)}]{kal10}
{Kale} R., {Dwarakanath} K.~S., 2010, \apj, 718, 939

\bibitem[{{Kale} \& {Dwarakanath}(2012)}]{kaldwa12}
{Kale} R., {Dwarakanath} K.~S., 2012, ApJ, 744, 46

\bibitem[{{Kale} {et~al}\mbox{.}(2012){Kale}, {Dwarakanath}, {Bagchi}, \&
  {Paul}}]{kal12}
{Kale} R., {Dwarakanath} K.~S., {Bagchi} J., {Paul} S., 2012, \mnras, 426, 1204

\bibitem[{{Kale} {et~al}\mbox{.}(2015){Kale}, {Venturi}, {Giacintucci},
  {Dallacasa}, {Cassano}, {Brunetti}, {Cuciti}, {Macario}, \&
  {Athreya}}]{kal15}
{Kale} R. {et~al.}, 2015, ArXiv e-prints

\bibitem[{{Kale} {et~al}\mbox{.}(2013){Kale}, {Venturi}, {Giacintucci},
  {Dallacasa}, {Cassano}, {Brunetti}, {Macario}, \& {Athreya}}]{kal13}
{Kale} R., {Venturi} T., {Giacintucci} S., {Dallacasa} D., {Cassano} R.,
  {Brunetti} G., {Macario} G., {Athreya} R., 2013, \aap, 557, A99

\bibitem[{{Kang}(2003)}]{2003JKAS...36..111K}
{Kang} H., 2003, Journal of Korean Astronomical Society, 36, 111

\bibitem[{{Kang} \& {Ryu}(2016)}]{2016arXiv160203278K}
{Kang} H., {Ryu} D., 2016, ArXiv e-prints

\bibitem[{{Keshet} \& {Loeb}(2010)}]{kes10}
{Keshet} U., {Loeb} A., 2010, \apj, 722, 737

\bibitem[{{Klamer}, {Subrahmanyan} \& {Hunstead}(2004){Klamer}, {Subrahmanyan},
  \& {Hunstead}}]{2004MNRAS.351..101K}
{Klamer} I., {Subrahmanyan} R., {Hunstead} R.~W., 2004, \mnras, 351, 101

\bibitem[{{Laing} \& {Bridle}(2002)}]{2002MNRAS.336..328L}
{Laing} R.~A., {Bridle} A.~H., 2002, \mnras, 336, 328

\bibitem[{{Laing} \& {Bridle}(2014)}]{2014MNRAS.437.3405L}
{Laing} R.~A., {Bridle} A.~H., 2014, \mnras, 437, 3405

\bibitem[{{Lal}(2009)}]{2009ASPC..407..157L}
{Lal} D.~V., 2009, in Astronomical Society of the Pacific Conference Series,
  Vol. 407, The Low-Frequency Radio Universe, {Saikia} D.~J., {Green} D.~A.,
  {Gupta} Y., {Venturi} T., eds., p. 157

\bibitem[{{Lal} {et~al}\mbox{.}(2010){Lal}, {Kraft}, {Forman}, {Hardcastle},
  {Jones}, {Nulsen}, {Evans}, {Croston}, \& {Lee}}]{2010ApJ...722.1735L}
{Lal} D.~V. {et~al.}, 2010, \apj, 722, 1735

\bibitem[{{Lal} {et~al}\mbox{.}(2013){Lal}, {Kraft}, {Randall}, {Forman},
  {Nulsen}, {Roediger}, {ZuHone}, {Hardcastle}, {Jones}, \&
  {Croston}}]{2013ApJ...764...83L}
{Lal} D.~V. {et~al.}, 2013, \apj, 764, 83

\bibitem[{{Lal} \& {Rao}(2004)}]{2004A&A...420..491L}
{Lal} D.~V., {Rao} A.~P., 2004, \aap, 420, 491

\bibitem[{{Large}, {Mathewson} \& {Haslam}(1959){Large}, {Mathewson}, \&
  {Haslam}}]{1959Natur.183.1663L}
{Large} M.~I., {Mathewson} D.~S., {Haslam} C.~G.~T., 1959, \nat, 183, 1663

\bibitem[{{Lindner} {et~al}\mbox{.}(2015){Lindner}, {Aguirre}, {Baker}, {Bond},
  {Crichton}, {Devlin}, {Essinger-Hileman}, {Gallardo}, {Gralla}, {Hilton},
  {Hincks}, {Huffenberger}, {Hughes}, {Infante}, {Lima}, {Marriage},
  {Menanteau}, {Niemack}, {Page}, {Schmitt}, {Sehgal}, {Sievers}, {Sif{\'o}n},
  {Staggs}, {Swetz}, {Wei{\ss}}, \& {Wollack}}]{2015ApJ...803...79L}
{Lindner} R.~R. {et~al.}, 2015, \apj, 803, 79

\bibitem[{{Loken} {et~al}\mbox{.}(1995){Loken}, {Roettiger}, {Burns}, \&
  {Norman}}]{1995ApJ...445...80L}
{Loken} C., {Roettiger} K., {Burns} J.~O., {Norman} M., 1995, \apj, 445, 80

\bibitem[{{Malu}, {Datta} \& {Sandhu}(2016){Malu}, {Datta}, \&
  {Sandhu}}]{2016Ap&SS.361..255M}
{Malu} S., {Datta} A., {Sandhu} P., 2016, ApSS, 361, 255

\bibitem[{{Malu} \& {Subrahmanyan}(2011)}]{2011JApA...32..541M}
{Malu} S.~S., {Subrahmanyan} R., 2011, Journal of Astrophysics and Astronomy,
  32, 541

\bibitem[{{Malu} {et~al}\mbox{.}(2010){Malu}, {Subrahmanyan}, {Wieringa}, \&
  {Narasimha}}]{2010arXiv1005.1394M}
{Malu} S.~S., {Subrahmanyan} R., {Wieringa} M., {Narasimha} D., 2010, ArXiv
  e-prints

\bibitem[{{Mazzotta} \& {Giacintucci}(2008)}]{maz08}
{Mazzotta} P., {Giacintucci} S., 2008, \apjl, 675, L9

\bibitem[{{Miley} {et~al}\mbox{.}(1972){Miley}, {Perola}, {van der Kruit}, \&
  {van der Laan}}]{1972Natur.237..269M}
{Miley} G.~K., {Perola} G.~C., {van der Kruit} P.~C., {van der Laan} H., 1972,
  \nat, 237, 269

\bibitem[{{Miller} {et~al}\mbox{.}(2013){Miller}, {Bonzini}, {Fomalont},
  {Kellermann}, {Mainieri}, {Padovani}, {Rosati}, {Tozzi}, \&
  {Vattakunnel}}]{2013ApJS..205...13M}
{Miller} N.~A. {et~al.}, 2013, \apjs, 205, 13

\bibitem[{{Miniati} {et~al}\mbox{.}(2000){Miniati}, {Ryu}, {Kang}, {Jones},
  {Cen}, \& {Ostriker}}]{min00}
{Miniati} F., {Ryu} D., {Kang} H., {Jones} T.~W., {Cen} R., {Ostriker} J.~P.,
  2000, ApJ, 542, 608

\bibitem[{{Mitchell} \& {Culhane}(1977)}]{1977MNRAS.178P..75M}
{Mitchell} R.~J., {Culhane} J.~L., 1977, \mnras, 178, 75P

\bibitem[{{Norris} {et~al}\mbox{.}(2011){Norris}, {Hopkins}, {Afonso}, {Brown},
  {Condon}, {Dunne}, {Feain}, {Hollow}, {Jarvis}, {Johnston-Hollitt}, {Lenc},
  {Middelberg}, {Padovani}, {Prandoni}, {Rudnick}, {Seymour}, {Umana},
  {Andernach}, {Alexander}, {Appleton}, {Bacon}, {Banfield}, {Becker}, {Brown},
  {Ciliegi}, {Jackson}, {Eales}, {Edge}, {Gaensler}, {Giovannini}, {Hales},
  {Hancock}, {Huynh}, {Ibar}, {Ivison}, {Kennicutt}, {Kimball}, {Koekemoer},
  {Koribalski}, {L{\'o}pez-S{\'a}nchez}, {Mao}, {Murphy}, {Messias},
  {Pimbblet}, {Raccanelli}, {Randall}, {Reiprich}, {Roseboom},
  {R{\"o}ttgering}, {Saikia}, {Sharp}, {Slee}, {Smail}, {Thompson}, {Urquhart},
  {Wall}, \& {Zhao}}]{2011PASA...28..215N}
{Norris} R.~P. {et~al.}, 2011, \pasa, 28, 215

\bibitem[{{Ogrean} {et~al}\mbox{.}(2014){Ogrean}, {Br{\"u}ggen}, {van Weeren},
  {Burgmeier}, \& {Simionescu}}]{ogr14}
{Ogrean} G.~A., {Br{\"u}ggen} M., {van Weeren} R.~J., {Burgmeier} A.,
  {Simionescu} A., 2014, \mnras, 443, 2463

\bibitem[{{Owen} {et~al}\mbox{.}(2014){Owen}, {Rudnick}, {Eilek}, {Rau},
  {Bhatnagar}, \& {Kogan}}]{owe14}
{Owen} F.~N., {Rudnick} L., {Eilek} J., {Rau} U., {Bhatnagar} S., {Kogan} L.,
  2014, \apj, 794, 24

\bibitem[{{Padmanabhan}(2000)}]{2000thas.book.....P}
{Padmanabhan} T., 2000, {Theoretical Astrophysics - Volume 1, Astrophysical
  Processes}. p. 622

\bibitem[{{Pandey-Pommier} {et~al}\mbox{.}(2013){Pandey-Pommier}, {Richard},
  {Combes}, {Dwarakanath}, {Guiderdoni}, {Ferrari}, {Sirothia}, \&
  {Narasimha}}]{2013A&A...557A.117P}
{Pandey-Pommier} M., {Richard} J., {Combes} F., {Dwarakanath} K.~S.,
  {Guiderdoni} B., {Ferrari} C., {Sirothia} S., {Narasimha} D., 2013, \aap,
  557, A117

\bibitem[{{Pandey-Pommier} {et~al}\mbox{.}(2015){Pandey-Pommier}, {van Weeren},
  {Ogrean}, {Combes}, {Johnston-Hollitt}, {Richard}, {Bagchi}, {Guiderdoni},
  {Jacob}, {Dwarakanath}, {Narasimha}, {Edge}, {Ebeling}, {Clarke}, \&
  {Mroczkowski}}]{2015sf2a.conf..247P}
{Pandey-Pommier} M. {et~al.}, 2015, in SF2A-2015: Proceedings of the Annual
  meeting of the French Society of Astronomy and Astrophysics, {Martins} F.,
  {Boissier} S., {Buat} V., {Cambr{\'e}sy} L., {Petit} P., eds., pp. 247--252

\bibitem[{{Parekh} {et~al}\mbox{.}(2016){Parekh}, {Dwarakanath}, {Kale}, \&
  {Intema}}]{2016arXiv160802796P}
{Parekh} V., {Dwarakanath} K.~S., {Kale} R., {Intema} H., 2016, ArXiv e-prints

\bibitem[{{Paul} {et~al}\mbox{.}(2011){Paul}, {Iapichino}, {Miniati}, {Bagchi},
  \& {Mannheim}}]{pau11}
{Paul} S., {Iapichino} L., {Miniati} F., {Bagchi} J., {Mannheim} K., 2011, ApJ,
  726, 17

\bibitem[{{Perucho}(2012)}]{2012IJMPS...8..241P}
{Perucho} M., 2012, International Journal of Modern Physics Conference Series,
  8, 241

\bibitem[{{Perucho}, {Quilis} \& {Mart{\'{\i}}}(2011){Perucho}, {Quilis}, \&
  {Mart{\'{\i}}}}]{2011ApJ...743...42P}
{Perucho} M., {Quilis} V., {Mart{\'{\i}}} J.-M., 2011, \apj, 743, 42

\bibitem[{{Perucho}, {Quilis} \& {Mart{\'{\i}}}(2012){Perucho}, {Quilis}, \&
  {Mart{\'{\i}}}}]{2012ASPC..459..155P}
{Perucho} M., {Quilis} V., {Mart{\'{\i}}} J.-M., 2012, in Astronomical Society
  of the Pacific Conference Series, Vol. 459, Numerical Modeling of Space
  Plasma Slows (ASTRONUM 2011), {Pogorelov} N.~V., {Font} J.~A., {Audit} E.,
  {Zank} G.~P., eds., p. 155

\bibitem[{{Petrosian}(2001)}]{pet01}
{Petrosian} V., 2001, \apj, 557, 560

\bibitem[{{Petrosian} \& {East}(2008)}]{pet08}
{Petrosian} V., {East} W.~E., 2008, \apj, 682, 175

\bibitem[{{Pfrommer} \& {En{\ss}lin}(2004)}]{pfr04}
{Pfrommer} C., {En{\ss}lin} T.~A., 2004, \aap, 413, 17

\bibitem[{{Pizzo} {et~al}\mbox{.}(2011){Pizzo}, {de Bruyn}, {Bernardi}, \&
  {Brentjens}}]{2011A&A...525A.104P}
{Pizzo} R.~F., {de Bruyn} A.~G., {Bernardi} G., {Brentjens} M.~A., 2011, \aap,
  525, A104

\bibitem[{{Planck Collaboration} {et~al}\mbox{.}(2014){Planck Collaboration},
  {Ade}, {Aghanim}, {Armitage-Caplan}, {Arnaud}, {Ashdown}, {Atrio-Barandela},
  {Aumont}, {Aussel}, {Baccigalupi}, \& et~al.}]{2014A&A...571A..29P}
{Planck Collaboration} {et~al.}, 2014, \aap, 571, A29

\bibitem[{{Planck Collaboration} {et~al}\mbox{.}(2013){Planck Collaboration},
  {Ade}, {Aghanim}, {Arnaud}, {Ashdown}, {Atrio-Barandela}, {Aumont},
  {Baccigalupi}, {Balbi}, {Banday}, \& et~al.}]{2013A&A...550A.134P}
{Planck Collaboration} {et~al.}, 2013, \aap, 550, A134

\bibitem[{{Planck Collaboration} {et~al}\mbox{.}(2011){Planck Collaboration},
  {Ade}, {Aghanim}, {Arnaud}, {Ashdown}, {Aumont}, {Baccigalupi}, {Balbi},
  {Banday}, {Barreiro}, \& et~al.}]{2011A&A...536A...8P}
{Planck Collaboration} {et~al.}, 2011, \aap, 536, A8

\bibitem[{{Raychaudhury} {et~al}\mbox{.}(1991){Raychaudhury}, {Fabian}, {Edge},
  {Jones}, \& {Forman}}]{1991MNRAS.248..101R}
{Raychaudhury} S., {Fabian} A.~C., {Edge} A.~C., {Jones} C., {Forman} W., 1991,
  \mnras, 248, 101

\bibitem[{{Reichardt} {et~al}\mbox{.}(2013){Reichardt}, {Stalder}, {Bleem},
  {Montroy}, {Aird}, {Andersson}, {Armstrong}, {Ashby}, {Bautz}, {Bayliss},
  {Bazin}, {Benson}, {Brodwin}, {Carlstrom}, {Chang}, {Cho}, {Clocchiatti},
  {Crawford}, {Crites}, {de Haan}, {Desai}, {Dobbs}, {Dudley}, {Foley},
  {Forman}, {George}, {Gladders}, {Gonzalez}, {Halverson}, {Harrington},
  {High}, {Holder}, {Holzapfel}, {Hoover}, {Hrubes}, {Jones}, {Joy}, {Keisler},
  {Knox}, {Lee}, {Leitch}, {Liu}, {Lueker}, {Luong-Van}, {Mantz}, {Marrone},
  {McDonald}, {McMahon}, {Mehl}, {Meyer}, {Mocanu}, {Mohr}, {Murray}, {Natoli},
  {Padin}, {Plagge}, {Pryke}, {Rest}, {Ruel}, {Ruhl}, {Saliwanchik}, {Saro},
  {Sayre}, {Schaffer}, {Shaw}, {Shirokoff}, {Song}, {Spieler}, {Staniszewski},
  {Stark}, {Story}, {Stubbs}, {{\v S}uhada}, {van Engelen}, {Vanderlinde},
  {Vieira}, {Vikhlinin}, {Williamson}, {Zahn}, \&
  {Zenteno}}]{2013ApJ...763..127R}
{Reichardt} C.~L. {et~al.}, 2013, \apj, 763, 127

\bibitem[{{Rhee}, {Burns} \& {Kowalski}(1994){Rhee}, {Burns}, \&
  {Kowalski}}]{1994AJ....108.1137R}
{Rhee} G., {Burns} J.~O., {Kowalski} M.~P., 1994, \aj, 108, 1137

\bibitem[{{Roettiger}, {Burns} \& {Loken}(1996){Roettiger}, {Burns}, \&
  {Loken}}]{1996ApJ...473..651R}
{Roettiger} K., {Burns} J.~O., {Loken} C., 1996, \apj, 473, 651

\bibitem[{{Ryle} \& {Windram}(1968)}]{1968MNRAS.138....1R}
{Ryle} M., {Windram} M.~D., 1968, \mnras, 138, 1

\bibitem[{{Scaramella} {et~al}\mbox{.}(1989){Scaramella}, {Baiesi-Pillastrini},
  {Chincarini}, {Vettolani}, \& {Zamorani}}]{1989Natur.338..562S}
{Scaramella} R., {Baiesi-Pillastrini} G., {Chincarini} G., {Vettolani} G.,
  {Zamorani} G., 1989, \nat, 338, 562

\bibitem[{{Spergel} {et~al}\mbox{.}(2003){Spergel}, {Verde}, {Peiris},
  {Komatsu}, {Nolta}, {Bennett}, {Halpern}, {Hinshaw}, {Jarosik}, {Kogut},
  {Limon}, {Meyer}, {Page}, {Tucker}, {Weiland}, {Wollack}, \&
  {Wright}}]{2003ApJS..148..175S}
{Spergel} D.~N. {et~al.}, 2003, \apjs, 148, 175

\bibitem[{{Springel} {et~al}\mbox{.}(2005){Springel}, {White}, {Jenkins},
  {Frenk}, {Yoshida}, {Gao}, {Navarro}, {Thacker}, {Croton}, {Helly},
  {Peacock}, {Cole}, {Thomas}, {Couchman}, {Evrard}, {Colberg}, \&
  {Pearce}}]{2005Natur.435..629S}
{Springel} V. {et~al.}, 2005, \nat, 435, 629

\bibitem[{{Stroe} {et~al}\mbox{.}(2016){Stroe}, {Shimwell}, {Rumsey}, {van
  Weeren}, {Kierdorf}, {Donnert}, {Jones}, {R{\"o}ttgering}, {Hoeft},
  {Rodr{\'{\i}}guez-Gonz{\'a}lvez}, {Harwood}, \&
  {Saunders}}]{2016MNRAS.455.2402S}
{Stroe} A. {et~al.}, 2016, \mnras, 455, 2402

\bibitem[{{Subramanian}, {Shukurov} \& {Haugen}(2006){Subramanian}, {Shukurov},
  \& {Haugen}}]{sub06}
{Subramanian} K., {Shukurov} A., {Haugen} N.~E.~L., 2006, \mnras, 366, 1437

\bibitem[{{Sunyaev} \& {Zeldovich}(1972)}]{1972CoASP...4..173S}
{Sunyaev} R.~A., {Zeldovich} Y.~B., 1972, Comments on Astrophysics and Space
  Physics, 4, 173

\bibitem[{{Trasatti} {et~al}\mbox{.}(2015){Trasatti}, {Akamatsu}, {Lovisari},
  {Klein}, {Bonafede}, {Br{\"u}ggen}, {Dallacasa}, \& {Clarke}}]{tra15}
{Trasatti} M., {Akamatsu} H., {Lovisari} L., {Klein} U., {Bonafede} A.,
  {Br{\"u}ggen} M., {Dallacasa} D., {Clarke} T., 2015, \aap, 575, A45

\bibitem[{{van Weeren} {et~al}\mbox{.}(2011){van Weeren}, {Br{\"u}ggen},
  {R{\"o}ttgering}, {Hoeft}, {Nuza}, \& {Intema}}]{wee11b}
{van Weeren} R.~J., {Br{\"u}ggen} M., {R{\"o}ttgering} H.~J.~A., {Hoeft} M.,
  {Nuza} S.~E., {Intema} H.~T., 2011, A\&A, 533, A35

\bibitem[{{van Weeren} {et~al}\mbox{.}(2014{\natexlab{a}}){van Weeren},
  {Intema}, {Lal}, {Andrade-Santos}, {Br{\"u}ggen}, {de Gasperin}, {Forman},
  {Hoeft}, {Jones}, {Nuza}, {R{\"o}ttgering}, \& {Stroe}}]{2014ApJ...786L..17V}
{van Weeren} R.~J. {et~al.}, 2014{\natexlab{a}}, \apjl, 786, L17

\bibitem[{{van Weeren} {et~al}\mbox{.}(2014{\natexlab{b}}){van Weeren},
  {Intema}, {Lal}, {Bonafede}, {Jones}, {Forman}, {R{\"o}ttgering},
  {Br{\"u}ggen}, {Stroe}, {Hoeft}, {Nuza}, \& {de
  Gasperin}}]{2014ApJ...781L..32V}
{van Weeren} R.~J. {et~al.}, 2014{\natexlab{b}}, \apjl, 781, L32

\bibitem[{{van Weeren} {et~al}\mbox{.}(2009){van Weeren}, {R{\"o}ttgering},
  {Bagchi}, {Raychaudhury}, {Intema}, {Miniati}, {En{\ss}lin}, {Markevitch}, \&
  {Erben}}]{wee09}
{van Weeren} R.~J. {et~al.}, 2009, A\&A, 506, 1083

\bibitem[{{van Weeren} {et~al}\mbox{.}(2010){van Weeren}, {R{\"o}ttgering},
  {Br{\"u}ggen}, \& {Hoeft}}]{wee10}
{van Weeren} R.~J., {R{\"o}ttgering} H.~J.~A., {Br{\"u}ggen} M., {Hoeft} M.,
  2010, Science, 330, 347

\bibitem[{{Vazza} \& {Br{\"u}ggen}(2014)}]{vaz14}
{Vazza} F., {Br{\"u}ggen} M., 2014, \mnras, 437, 2291

\bibitem[{{Vazza} {et~al}\mbox{.}(2015{\natexlab{a}}){Vazza}, {Ferrari},
  {Bonafede}, {Br{\"u}ggen}, {Gheller}, {Braun}, \&
  {Brown}}]{2015aska.confE..97V}
{Vazza} F., {Ferrari} C., {Bonafede} A., {Br{\"u}ggen} M., {Gheller} C.,
  {Braun} R., {Brown} S., 2015{\natexlab{a}}, Advancing Astrophysics with the
  Square Kilometre Array (AASKA14), 97

\bibitem[{{Vazza} {et~al}\mbox{.}(2015{\natexlab{b}}){Vazza}, {Ferrari},
  {Br{\"u}ggen}, {Bonafede}, {Gheller}, \& {Wang}}]{2015A&A...580A.119V}
{Vazza} F., {Ferrari} C., {Br{\"u}ggen} M., {Bonafede} A., {Gheller} C., {Wang}
  P., 2015{\natexlab{b}}, \aap, 580, A119

\bibitem[{{Venkatesan} {et~al}\mbox{.}(1994){Venkatesan}, {Batuski}, {Hanisch},
  \& {Burns}}]{1994ApJ...436...67V}
{Venkatesan} T.~C.~A., {Batuski} D.~J., {Hanisch} R.~J., {Burns} J.~O., 1994,
  \apj, 436, 67

\bibitem[{{Venturi} {et~al}\mbox{.}(2000){Venturi}, {Bardelli}, {Morganti}, \&
  {Hunstead}}]{2000MNRAS.314..594V}
{Venturi} T., {Bardelli} S., {Morganti} R., {Hunstead} R.~W., 2000, \mnras,
  314, 594

\bibitem[{{Venturi} {et~al}\mbox{.}(2007){Venturi}, {Giacintucci}, {Brunetti},
  {Cassano}, {Bardelli}, {Dallacasa}, \& {Setti}}]{ven07}
{Venturi} T., {Giacintucci} S., {Brunetti} G., {Cassano} R., {Bardelli} S.,
  {Dallacasa} D., {Setti} G., 2007, \aap, 463, 937

\bibitem[{{Venturi} {et~al}\mbox{.}(2008){Venturi}, {Giacintucci}, {Dallacasa},
  {Cassano}, {Brunetti}, {Bardelli}, \& {Setti}}]{ven08}
{Venturi} T., {Giacintucci} S., {Dallacasa} D., {Cassano} R., {Brunetti} G.,
  {Bardelli} S., {Setti} G., 2008, A\&A, 484, 327

\bibitem[{{Willson}(1970)}]{1970MNRAS.151....1W}
{Willson} M.~A.~G., 1970, \mnras, 151, 1

\bibitem[{{Zel'dovich}(1970)}]{1970A&A.....5...84Z}
{Zel'dovich} Y.~B., 1970, \aap, 5, 84

\bibitem[{{ZuHone} {et~al}\mbox{.}(2014){ZuHone}, {Brunetti}, {Giacintucci}, \&
  {Markevitch}}]{zuh14}
{ZuHone} J., {Brunetti} G., {Giacintucci} S., {Markevitch} M., 2014, ArXiv
  e-prints

\bibitem[{{ZuHone} {et~al}\mbox{.}(2013){ZuHone}, {Markevitch}, {Brunetti}, \&
  {Giacintucci}}]{zuh13}
{ZuHone} J.~A., {Markevitch} M., {Brunetti} G., {Giacintucci} S., 2013, \apj,
  762, 78

\end{thebibliography}

\end{document}